 \documentclass[aip,jcp,reprint,amsmath,amssymb,showpacs]{revtex4-1}%
\usepackage{hyperref}
\usepackage{graphicx}
 \usepackage{amsmath}
\usepackage{graphicx}%
\usepackage{amsfonts}%
 \usepackage{amssymb}
\usepackage{color}
\usepackage{setspace}
\usepackage{textcomp}

\begin{document}

\title{Scaling of selectivity in uniformly charged nanopores through a modified Dukhin number for 1:1 electrolytes } 
\author{Zs\'ofia Sarkadi}
\author{D\'avid Fertig}
\author{M\'onika Valisk\'o}
\author{Dezs\H{o} Boda}\email[Author for correspondence:]{boda@almos.vein.hu}

% \author{M\'onika Valisk\'o$^{1}$, Bart\l{}omiej Matejczyk$^{2}$, Zolt\'an Hat\'o$^{1}$, Tam\'as Krist\'of$^{1}$, Eszter M\'adai$^{1,3}$, D\'avid Fertig$^{1}$,  Dezs\H{o} Boda$^{1}$}
% \email[Author for correspondence:]{valisko@almos.vein.hu}
\affiliation{$^{1}$Department of Physical Chemistry, University of Pannonia, P.O. Box 158, H-8201 Veszpr\'em, Hungary}
% \affiliation{$^{4}$Department of Physiology and Biophysics, Rush University Medical Center, Chicago, IL 60612, USA}
\date{\today}

% Keywords: activity coefficient, solvation, Grand Canonical Monte Carlo, Born energy

\begin{abstract}
We show that a modified version of the Dukhin number is an appropriate scaling parameter for the ionic selectivity of uniformly charged nanopores.
The modified Dukhin number is an unambiguous function of the variables $\sigma$ (surface charge), $R$ (pore radius), and $c$ (salt concentration), and defined as $\mathrm{mDu}=|\sigma|/e(R/\lambda)$, where $\lambda$ is the screening length of the electrolyte carrying the $c$ dependence ($\lambda\sim c^{-1/2}$).
Scaling means that the device function (selectivity) is a smooth and (in this case) monotonic function of mDu.
The original Dukhin number defined as $\mathrm{Du}=|\sigma|/eRc$ ($c^{-1}$ dependence) was introduced to indicate whether the surface or the volume conduction is dominant in the pore.
The modified version satisfies scaling and characterizes selectivity in the intermediate regime, where both surface and bulk conductions are present and the pore is neither perfectly selective, nor perfectly non-selective.
Our modeling study using the Local Equilibrium Monte Carlo method and the Poisson-Nernst-Planck theory  provides the radial flux profiles from which the radial selectivity profile can be computed.
These profiles show in which region of the nanopore the surface or the volume conduction dominates for a given combination of the variables $\sigma$, $R$, and $c$.
We show that the inflection point of the scaling curve may be used to characterize the transition point between the surface and volume conductions.

% \begin{center}
% \includegraphics*[width=0.4\textwidth]{figs/toc}\\
% TOC figure
% \end{center}
\end{abstract}

% \pacs{02.70.-c, 02.70.Uu, 05.10.Ln}

\maketitle

%#######################################################################
%#######################################################################

%#######################################################################
%#######################################################################
\section{Introduction}
\label{sec:intro}

When the function of a device is determined by a few well-defined variables, $a_{1}, a_{2}, \dots$, it is often possible to group them into a composite parameter, $\xi$, that also determines the device's behavior.
% We call this phenonemon scaling and we call this composite parameter scaling parameter.
This scaling parameter is a well-defined function of the independent variables: $\xi=\xi(a_{1}, a_{2}, \dots)$.
The device function, $F$, is an obervable property of the device.
Scaling of the device function means that $F$ is a smooth unambiguous function of the scaling parameter: $F=f\left[ \xi(a_{1}, a_{2}, \dots) \right]$.

Nanopores are located in a membrane and connect two bath electrolytes.
They facilitate the controlled movement of ions from one side of the membrane to the other side.
Tunable input parameters of this device are the radius of the pore, $R$, voltage applied across the membrane, $U$, the surface charge on the wall of the nanopore, $\sigma$, ionic concentrations in the baths, $c$, and properties of the electrolyte, like ionic valences, $z_{i}$, for example ($i$ indexes the ionic species).
The measurable output parameters are the currents carried by the ions, $I_{i}$.

The structural properties of the nanopore (charge pattern and geometry) determine what is a practical choice for the device function.
In the case of a uniformly charged nanopore ($\sigma<0$ is constant) studied here (cation) selectivity defined as 
\begin{equation}
 S_{+}=\frac{I_{+}}{I_{+}+I_-} 
 \label{eq:sel}
\end{equation} 
is an appropriate device function.
It is an unambiguous function of the currents and is well-measurable via the reversal potential.
For 1:1 electrolytes, if this number is $\approx 0.5$, the pore is non-selective, while if it is $1$, the pore is perfectly cation selective.
Note that $S_-=1-S_{+}$.

In a previous paper,~\cite{fertig_jpcc_2019} we introduced the scaling parameter
\begin{equation}
 \xi = \dfrac{R}{\lambda \sqrt{z_{+}|z_-|}} ,
 \label{eq:xi}
\end{equation}
where $\lambda$ is a characteristic screening length of the electrolyte for which the Debye length is an obvious choice:
\begin{equation}
\lambda_{\mathrm{D}} = 
% \left( \dfrac{e^{2}}{\epsilon_{0}\epsilon kT} \sum_{i} z_{i}^{2}c_{i} \right)^{-1/2} = 
\left( \dfrac{ c e^{2}}{\epsilon_{0}\epsilon kT} \sum_{i} z_{i}^{2}\nu_{i} \right)^{-1/2} ,
\label{eq:lambdaD}
\end{equation} 
where $k$ is Boltzmann's constant, $T$ is the absolute temperature (it is $298.15$ K in this work), $e$ is the elementary charge, $\nu_{i}$ is the stoichiometric coefficent of ionic species $i$, $c$ is the salt concentration ($c_{i}=\nu_{i}c$ is the bath concentration of species $i$), $\epsilon$ is the dielectric constant of the solvent (it is $78.45$ in this work), and $\epsilon_0$ is the permittivity of vacuum.

The Debye length depends on the square root of the concentration, $\lambda_{\mathrm{D}}\sim 1{/}\sqrt{c}$, and characterizes the width of the double layer (DL) formed at the charged wall of the nanopore. 
Another choice \cite{fertig_jpcc_2019} for the screening length is the one obtained from the Mean Spherical Approximation (MSA)  \cite{blum_mp_1975,blum_jcp_1977,nonner_bj_2000} denoted by $\lambda_{\mathrm{MSA}}$. 

We showed \cite{fertig_jpcc_2019} that in a pore with a bipolar charge pattern (positive/negative) an obvious device function is the ratio of currents at forward- and reverse-biased values of voltage (rectification) and it scales with $\xi$.
In another paper, \cite{madai_pccp_2018} we showed that in a pore with a transistor-like charge pattern (positive/negative/positive) the ratio of currents in open and closed states (switching) scales with $R/\lambda_{\mathrm{D}}$ for a 1:1 electrolyte.

In these papers,~\cite{fertig_jpcc_2019,madai_pccp_2018} we studied nanopores with dimensions small enough that regions in the pore with very small ionic concentrations (depletion zones) form and determine device behavior.
% The nanopore is short, so the pore's behavior can be tuned by the depletion zones effieciently.
The regions of a nanopore along $z$-axis can be considered as resistors connected in series.
If the resistance of one segment is large due to the low concentration of an ionic species there (depletion zone), then the resistance is large for the whole pore for that ionic species.
Rectification of a bipolar nanopore, for example, is based on the fact that the depletion zones are deeper at one sign of the voltage than at the opposite sign.
A depletion zone for a given ionic species in a given region may form if that ionic species is the coion with respect to the surface charge in that region.

Formation of DLs, their overlap, and the resulting exclusion of coions are also the key factor determining selectivity.
If the DL at the wall is wide compared to the pore radius ($\lambda \gg R$), the DLs overlap and a bulk electrolyte is not formed around the centerline. 
The concentration of the coion is small not only close to the surface but also at the centerline (Fig.~\ref{fig1}).
The coions are depleted, counterions are in excess, and the surface conduction dominates.

If the width of the DL is small compared to the pore radius ($\lambda \ll R$), the DL is restricted to the region close to the surface and the region along the centerline contains enough coions so they can carry current in that region.
In the region where both coions and counterions are present, volume conduction dominates.

Between the two limiting cases both surface and volume conductions are present and our negative pore is not perfectly cation selective ($S_{+}$ is in between $0.5$ and $1$ for the 1:1 electrolyte consedered here).
% Double layer overlap is determined by the $R/ \lambda_{\mathrm{D}}$ ratio.

The calculations reported in Ref.~\onlinecite{fertig_jpcc_2019} were performed for a fixed value of the surface charge $\sigma {\pm} 1$ $e$/nm$^{2}$, so the $\xi$ scaling parameter does not contain $\sigma$.
Here, we pursue a scaling parameter that also includes $\sigma$ in addition to $R$ and $c$.

%%%%%%%%%%%%%%%%%%%%%%%%%%%%%%%%%%%%%%%%%%%%%%%%%%%%%%%%%%%%%%%%%%%%%%%%%%%%%%%%%%%%%%%%%%%%%%%%%%%%%%%%%%%%%%%%%%%%%%%%%%%%%%%%%%%%%%%%%%%%%%%%%%%%%%%%%%%%%%%%%%%%%%%%%%%%%%%%%%%%%%%%%%%%%%%%%%%%%%%%%%%%%%%%%%%%%%%%%%%%%%%%%%%%%%%%%%%%%%%%%%%%%%%%%%%%%%%%%%%%%%%%%%%%%%%%%%%%%%%%%%%%%%%%%%%%%%%%%%%%%%%%%%%%%%%%%%%%%%%%%

\section{Modified Dukhin number as scaling parameter}

% One candidate jumps in mind right away.

% The conductances in the formula were either measured or calculated.

Because there is a monotonic relationship between ionic selectivity and the ratio of surface and volume conductances that are present in the nanopore at the same time, the Dukhin number \cite{bazant_pre_2004,chu_pre_2006,bocquet_chemsocrev_2010} defined as
\begin{equation}
 \mathrm{Du} \equiv \frac{|\sigma|}{eRc} 
 \label{eq:dukhin}
\end{equation} 
offers itself to be the scaling parameter we are after. 
This definition of Du contains exactly those variables that we want to see in our scaling parameter, $\sigma$, $R$, and $c$. 

Du was originally introduced by Bikerman \cite{bikerman_1940} to characterize the ratio of the surface and volume conductances focusing on electrokinetic phenomena.
Later, Dukhin adopted the idea (see Ref.~\onlinecite{dukhin_advcollsci_1993} and references therein) to study electrophoretic phenomena.
Although the name Bikerman number also occurs in the literature, Lyklema introduced the name Dukhin number to salute Dukhin.~\cite{lyklema_book_1995}
The Dukhin number (and a characteristic length called Dukhin length, $l_{\mathrm{Du}}=\mathrm{Du}\, R$) was used in several modeling studies describing nanopores, more specifically, when volume and surface transport processes competed inside the pore.~\cite{bazant_pre_2004,chu_pre_2006,khair_jfm_2008,das_langmuir_2010,bocquet_chemsocrev_2010,zangle_csr_2010,lee_nanolett_2012,yeh_ijc_2014,ma_acssens_2017,xiong_scc_2019,poggioli_jpcb_2019,dalcengio_jcp_2019,kavokine_annualrev_2020,noh_acsnano_2020}

The definition in Eq.~\ref{eq:dukhin} is in agreement with the traditional definition of Bikerman, because it can be computed from the ratio of the surface excess of the cations, $|\sigma| 2\pi R H$ ($H$ is the length of the pore) assuming perfect exclusion of the anions, and the number of charge carriers assuming a bulk electrolyte in the pore, $2c R^{2}\pi H$ (the factor $2$ is needed because both cations and anions carry current). 

So, we have a parameter that seems to be appropriate for our purposes, we just need to check whether it works as a scaling parameter. 
We will see in Section \ref{sec:results} that it does not.

Therefore, we define a modified Dukhin number as
\begin{equation}
 \mathrm{mDu} \equiv \frac{|\sigma| }{eR/\lambda} .
 \label{eq:mDu}
\end{equation} 
As defined, mDu is not strictly a ``number,'' as it has units of m$^{-2}$.  
Below, we show that mDu is a scaling parameter for $S_+$.  
Therefore, it captures the essential physics of this device function.  
mDu can be multiplied by any length squared that does not depend on $R$ or $c$ (e.g., the Bjerrum length) to make it dimensionless, but this factor is irrelevant for the scaling so we will use the dimensional version throughout (except Section \ref{sec:inflexion}). 

Because $\lambda \sim 1/\sqrt{c}$, we should realize that
\begin{equation}
 \mathrm{mDu} \sim \frac{|\sigma|}{eR\sqrt{c}},
 \label{eq:mDu2}
\end{equation} 
namely, mDu corresponds to the original Dukhin number (Eq.~\ref{eq:dukhin}) with $\sqrt{c}$ in the denominator instead of $c$.
The importance of $\sqrt{c}$ is not a new idea.
Bikerman writes in his 1940 paper\cite{bikerman_1940}: ``At a constant $\zeta$ potential the surface conductivity $\chi$ is nearly proportional to the square root of the ionic concentration ($C^{\frac{1}{2}}$) whilst [the volume conductivity] $\kappa$ in dilute solutions is nearly proportional to the concentration $C$ itself. Thus, $\chi/\kappa = \mathrm{const}.\; C^{-\frac{1}{2}} $.'' 

The original and modified Dukhin numbers behave the same way in the two limiting cases, at very low and very large concentrations ($\lambda {\gg} R$ and $\lambda{\ll} R$).
% , but they also contain .
In between the limiting cases, however, they behave differently.  
Both contain $\sigma$ as a parameter, so both can account for the fact that the surface charge produces the DLs, and, thus, selectivity.

The original Dukhin number is obtained from the relation of the two limiting cases.
In these limiting cases the screening length has no importance, because there is no DL in the case of perfect volume conductance, while there are no coions in the case of perfect surface conductance (perfect DL overlap), so the ratio of $\lambda$ to $R$ does not matter. 

The modified Dukhin number brings in the Debye length that, in our opinion, is vital in constructing a proper scaling parameter.
In the realm between the limiting cases DLs may just partially overlap, so the ratio of $\lambda$ to $R$ characterizing the extent of DL overlap matters. 
We will show in Subsection \ref{sec:res-scaling} that mDu is a proper scaling parameter in the intermediate domain. 

The ratio of the surface and volume conductions is also interesting, but they are consequences. 
Surface conduction dominates because the DLs overlap and not vice versa.
In the intermediate domain, the nanopore contains both a region where rather the surface conduction dominates (near the wall) and a region where rather volume conduction dominates (in the middle of the pore around the centerline). 

The boundary between these regions, however, is not sharp, so any definition of surface and volume conductions is arbitrary, while selectivity is a well-defined measurable quantity.
That is why we focus on selectivity as our main device function.
We also discuss surface and volume conductions in Subsection \ref{sec:res:radialsel}, however, by analyzing our simulation results that allow us to look into the black box and see what is going on at the molecular level.
Concentration and flux profiles reveal the existence of DL and bulk regions inside the pore.

We think that a parameter intended to characterize surface vs.\ volume conduction should satisfy scaling.
The purpose of this paper is to show, using computer simulations, that mDu is an appropriate scaling parameter for the selectivity of a uniformly charged pore. 

%%%%%%%%%%%%%%%%%%%%%%%%%%%%%%%%%%%%%%%%%%%%%%%%%%%%%%%%%%%%%%%%%%%%%%%%%%%%%%%%%%%%%%%%%%%%%%%%%%%%%%%%%%%%%%%%%%%%%%%%%%%%%%%%%%%%%%%%%%%%%%%%%%%%%%%%%%%%%%%%%%%%%%%%%%%%%%%%%%%%%%%%%%%%%%%%%%%%%%%%%%%%%%%%%%%%%%%%%%%%%%%%%%%%%%%%%%%%%%%%%%%%%%%%%%%%%%%%%%%%%%%%%%%%%%%%%%%%%%%%%%%%%%%%%%%%%%%%%%%%%%%%%%%%%%%%%%%%%%%%%

\section{Results}
\label{sec:results}

\subsection{Model of the nanopore}

A cylindrical nanopore of length $H$ and radius $R$ spans a membrane that separates two baths.
The cylindrical wall of the nanopore and the flat parallel walls confining the membrane are assumed to be hard. 
The $z$ dimension is the one perpendicular to the membrane along the pore.
Because the system has rotational symmetry about the $z$ axis, the other relevant coordinate is the radial one, $r$, that represents the distance from the $z$ axis.

The electrolyte is modeled in the implicit solvent framework, namely, the interaction potential between two hard-sphere ions is defined by Coulomb's law in a dielectric background:
\begin{equation}
 u_{ij}(r)=
 \left\lbrace 
\begin{array}{ll}
\infty & \quad \mathrm{if} \quad r<(d_{i}+d_{j})/2 \\
 \dfrac{1}{4\pi\epsilon_{0}\epsilon} \dfrac{z_{i}z_{j}e^{2}}{r} & \quad \mathrm{if} \quad r \geq (d_{i}+d_{j})/2\\
\end{array}
\right. 
\label{eq:uij}
\end{equation}
where $d_i$ is the diameter of ionic species $i$, and $r$ is the distance between two ions. 
Here, we consider only 1:1 electrolytes, namely, $z_{+}=1$ and $z_-=-1$.
For the ionic diameters, we use $d_{+}=d_-=0.3$ nm.
A uniform negative surface charge $\sigma$ is placed on the wall of the nanopore.

\begin{figure}[t!]
\begin{center}
\includegraphics*[width=0.49\textwidth]{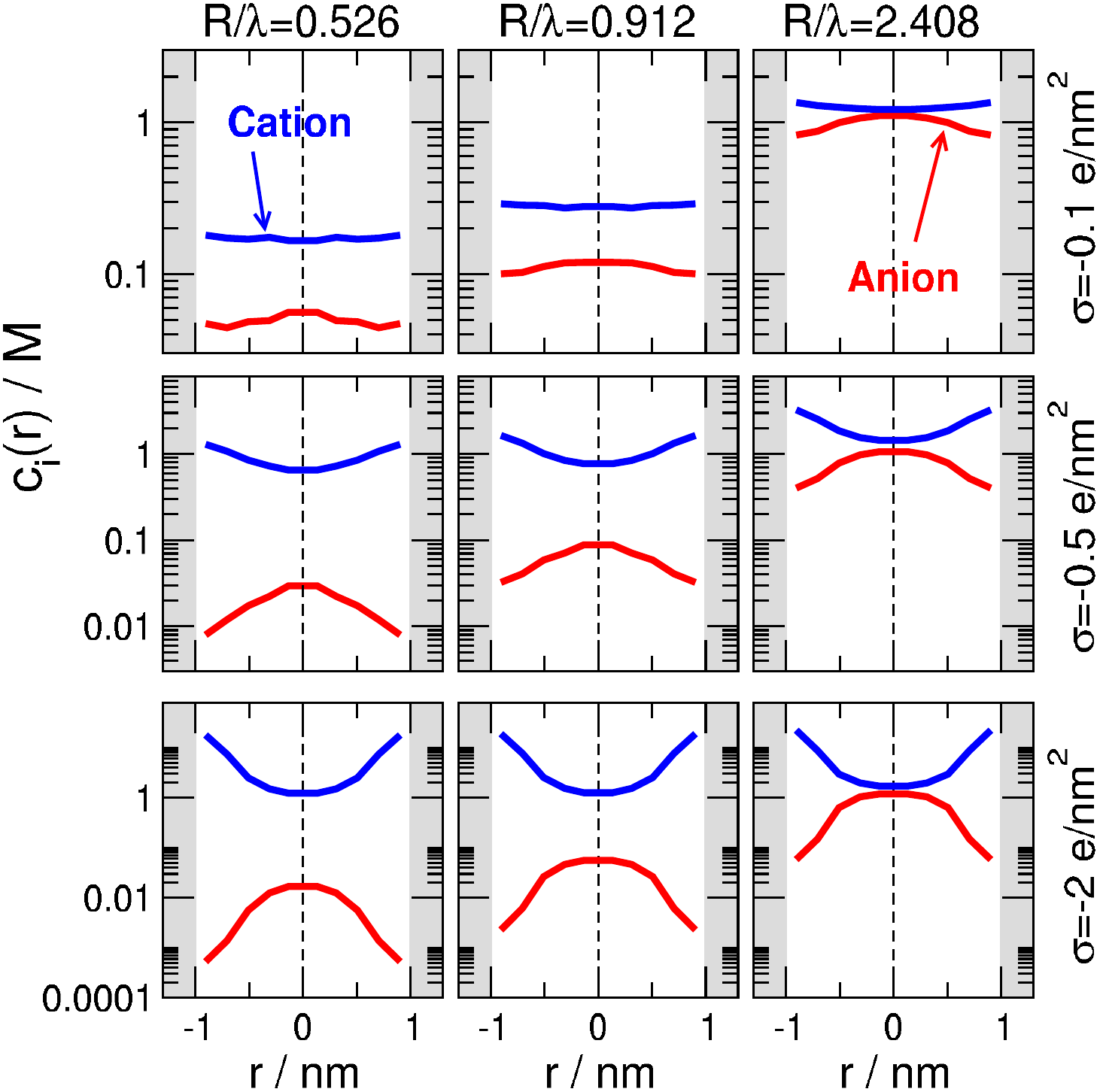}
\end{center}
\caption{Radial concentration profiles as obtained for various combinations of $\sigma$ and $R/\lambda$.
Different columns have different values of $R/\lambda$, while different rows have different values of $\sigma$.
In these cases $R=1$ nm. 
The concentrations that correspond to the $R/\lambda =0.526$, $0.912$, and $2.408$ values are $c=0.03$, $0.1$, and $1$ M, respectively.
Blue and red curves refer to the cations and the anions, respectively.
The curves are mirrored to $r=0$ for better visualization, while strictly speaking $r{\geq} 0$.
The applied method is NP+LEMC, so $\lambda=\lambda_{\mathrm{MSA}}$.
}
\label{fig1}
\end{figure}

The description of the methods with which we study this model is included in the Appendix.
Both methods are based on the Nernst-Planck (NP) transport equation (Eq.~\ref{eq:np}). 
In one method, we relate the concentration profile, $c_{i}(\mathbf{r})$, to the electrochemical potential profile, $\mu_{i}(\mathbf{r})$ with the Poisson-Boltzmann (PB) theory.
This continuum theory is known as the Poisson-Nernst-Planck (PNP) theory.

The other method is based on a Monte Carlo (MC) technique that is an adaptation of the Grand Canonical Monte Carlo (GCMC) method to a non-equilibrium situation, where $\mu_{i}(\mathbf{r})$ is not constant system-wide, so the system is not in global equilibrium, only in local equilibrium.
The method is called the Local Equilibrium Monte Carlo (LEMC) technique, while it is called NP+LEMC when we couple it to the NP equation.

Because the difference between the computational methods is that NP+LEMC includes all the ionic correlations present in the system and PNP is mean field, we use different screening lengths that reflect this difference.
So $\lambda= \lambda_{\mathrm{MSA}}$ if we use NP+LEMC, while $\lambda=\lambda_{\mathrm{D}}$ if we use PNP.

\begin{figure*}[t!]
\begin{center}
\includegraphics*[width=0.75\textwidth]{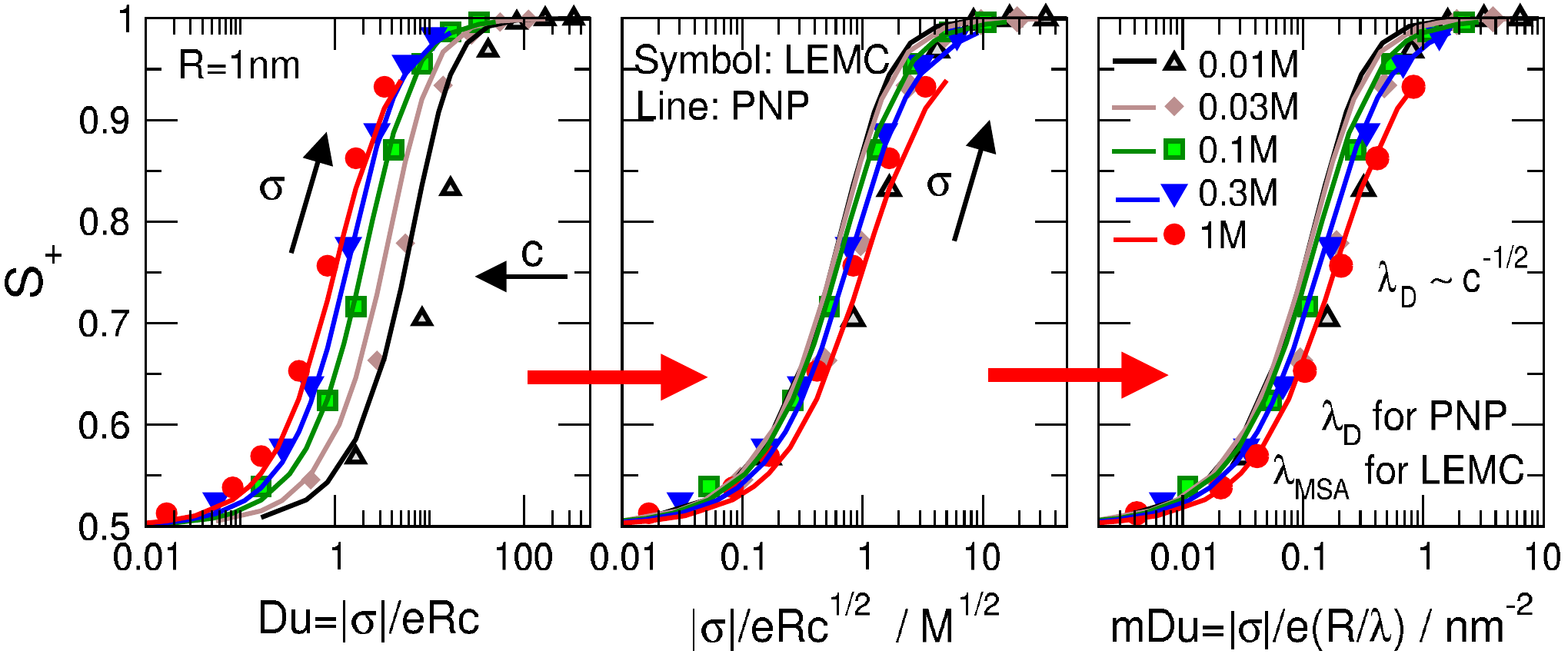}
\end{center}
\caption{The selectivity curves plotted against the Dukhin number, $\mathrm{Du}=|\sigma|/eRc$ (left panel), against the parameter in which there is $\sqrt{c}$ in place of $c$ in Du (middle panel), and against the modified Dukhin number, $\mathrm{mDu}=|\sigma|/e(R/\lambda )$ (right panel).
Symbols and lines refer to NP+LEMC and PNP results, respectively.
Different colors mean different concentrations, while points with a given color correspond to different surface charges.
The pore radius is constant, $R=1$ nm.
In the right panel, the MSA screening length ($\lambda_{\mathrm{MSA}}$) is used for the NP+LEMC data, while the Debye length ($\lambda_{\mathrm{D}}$) for the PNP data.
}
\label{fig2}
\end{figure*}

%%%%%%%%%%%%%%%%%%%%%%%%%%%%%%%%%%%%%%%%%%%%%%%%%%%%%%%%%%%%%%%%%%%%%%%%%%%%%%%%%%%%%%%%%%%%%%%%%%%%%%%%%%%%%%%%%%%%%%%%%%%%%%%%%%%%%%%%%%%%%%%%%%%%%%%%%%%%%%%%%%%%%%%%%%%%%%%%%%%%%%%%%%%%%%%%%%%%%%%%%%%%%%%%%%%%%%%%%%%%%%%%%%%%%%%%%%%%%%%%%%%%%%%%%%%%%%%%%%%%%%%%%%%%%%%%%%%%%%%%%%%%%%%%%%%%%%%%%%%%%%%%%%%%%%%%%%%%%%%%%

\subsection{Double layer formation and double layer overlap}

Figure \ref{fig1} illustrates the effect of the two relevant parameters ($\sigma$ and $R/\lambda$). 
The surface charge produces the separation of counterions and coions (DL formation), while the $R/ \lambda$ parameter accounts for the degree of overlap of the DL.
The figure shows radial concentration profiles to depict the interplay of these two effects. 

The columns correspond to different $R/\lambda$ values (a small $R/\lambda$ value means stronger DL overlap), while the rows correspond to different surface charges (a larger $|\sigma|$ value results in a stronger separation of cation and anion profiles).

As we go from left to right (increasing $R/\lambda$), the gap between the cation and anion profiles around the centerline $r\approx0$ decreases (note the log scale).
As we go from top to bottom (increasing $|\sigma|$), the gap between the cation and anion profiles near the wall ($r\approx R$) increases.

This implies that the two parameters have two relatively distinct effects.
The $R/\lambda$ parameter rather determines the behavior in the middle of the pore, while $\sigma$ rather determines the behavior in the DL.
It is common to express this distinction in terms of volume (or bulk) and surface conductions.
The DL region (if exists) is responsible for the surface conduction, while the bulk region in the middle (if exists) is responsible for the volume conduction.
With $R/\lambda$ and $\sigma$, therefore, we have two parameters with which we can tune the weights of the volume and surface conductions in the total conduction.
Later in this paper, we will provide a quantitative analysis for this.

%%%%%%%%%%%%%%%%%%%%%%%%%%%%%%%%%%%%%%%%%%%%%%%%%%%%%%%%%%%%%%%%%%%%%%%%%%%%%%%%%%%%%%%%%%%%%%%%%%%%%%%%%%%%%%%%%%%%%%%%%%%%%%%%%%%%%%%%%%%%%%%%%%%%%%%%%%%%%%%%%%%%%%%%%%%%%%%%%%%%%%%%%%%%%%%%%%%%%%%%%%%%%%%%%%%%%%%%%%%%%%%%%%%%%%%%%%%%%%%%%%%%%%%%%%%%%%%%%%%%%%%%%%%%%%%%%%%%%%%%%%%%%%%%%%%%%%%%%%%%%%%%%%%%%%%%%%%%%%%%%

\subsection{Scaling as a function of the modified Dukhin number}
\label{sec:res-scaling}

To check whether Du is an appropriate scaling parameter for our interests, we performed a large number of simulations (using both NP+LEMC and PNP) first for a fixed pore radius $R=1$ nm for varying values of $\sigma$ and $c$.
The results for our chosen device function, selectivity ($S_{+}$) are shown in the left panel of Fig.~\ref{fig2}.
The different curves are for different concentrations.
A given curve was obtained by varying surface charge.
The results show that Du is not appropriate for our purposes, because the curves do not coincide: they are spread along with the abscissa larger concentrations belonging to the smaller Du values.

Eq.~\ref{eq:mDu2} implies that the modified Dukhin number with $\lambda_{\mathrm{D}}$ in it is proportional to the original Dukhin number substituted $\sqrt{c}$ for $c$.
The results plotted as functions of $|\sigma|/eR\sqrt{c}$ are shown in the middle panel of Fig.~\ref{fig2}.
Now the curves nicely coincide, implying that the scaling factor we are looking for indeed depends on $\sqrt{c}$, and, thus, on $\lambda_{\mathrm{D}}$.

The $Rc^{1/2}$ product in the denominator is proportional to $R/\lambda_{\mathrm{D}}$, so it is conceptually advantageous to include the $R/ \lambda$ term in the new scaling parameter (Eq.~\ref{eq:mDu}) because it expresses the tendency of the DLs to overlap.

The right panel of Fig.~\ref{fig2} shows the results as functions of this new scaling parameter, mDu.
Apart from a rescaled abscissa, this panel is the same as the middle one with one notable difference that is hardly noticable with the naked eye.
This panel defines the screening length differently for the NP+LEMC and PNP data.
In the case of NP+LEMC we use the MSA screening length ($\lambda_{\mathrm{MSA}}$), while in the case of PNP we use the Debye length ($\lambda_{\mathrm{D}}$).
The difference between these is small especially using a logarithmic scale at the abscissa.
The difference, however, is larger at larger concentrations.

\begin{figure}[t]
\begin{center}
\includegraphics*[width=0.35\textwidth]{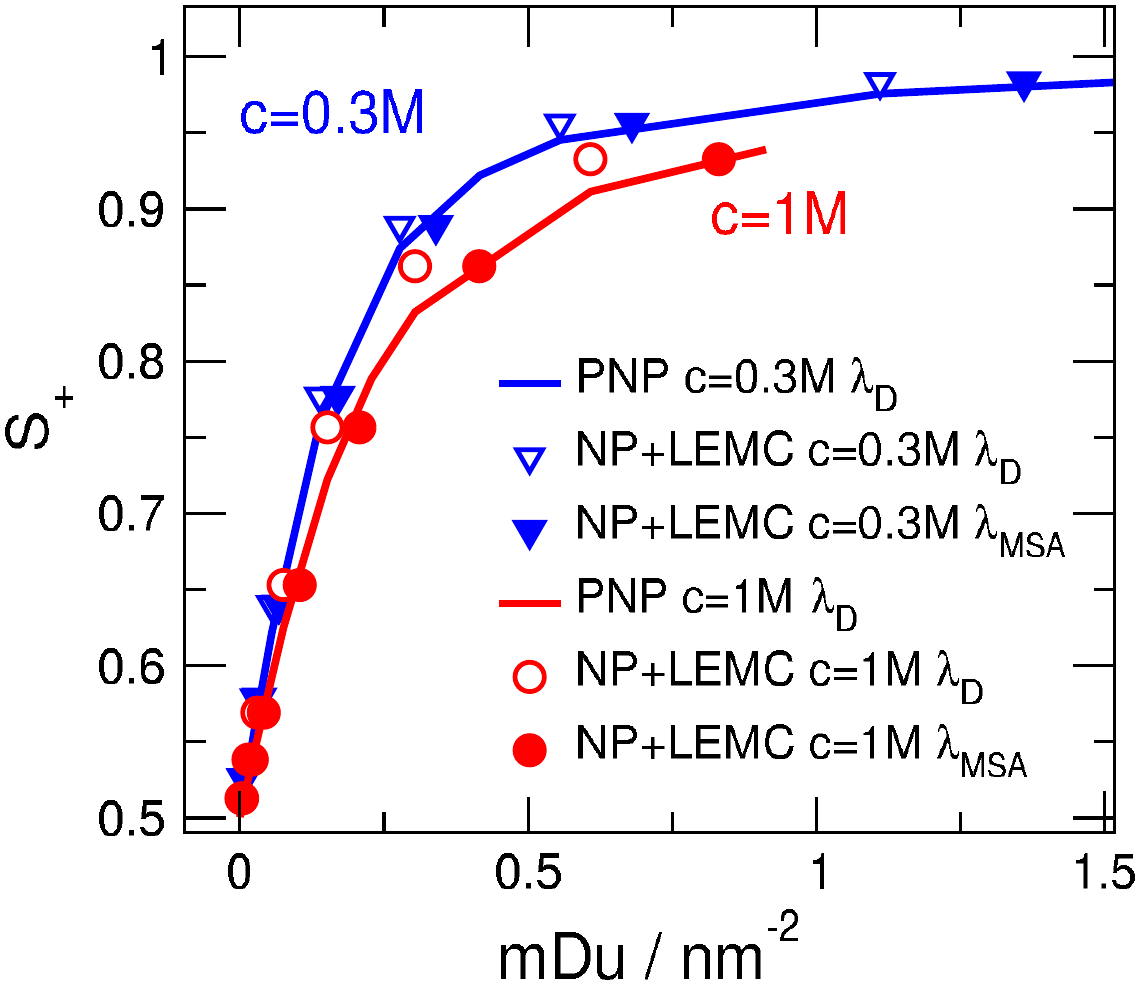}
\end{center}
\caption{The blue ($c=0.3$ M) and red ($c=1$ M) curves of Fig.~\ref{fig2} plotted on a linear scale of the abscissa (mDu).
The open and filled symbols show NP+LEMC values computed with $\lambda_{\mathrm{D}}$ and $\lambda_{\mathrm{MSA}}$, respectively.
The curves refer to PNP results using  $\lambda_{\mathrm{D}}$.
}
\label{fig3}
\end{figure}

This is better seen in Fig.~\ref{fig3} where only the data for $c=0.3$ and $1$ M are shown with a linear scale at the abscissa.
If one observes, for example, the red symbols and curves for $c=1$ M, one can see that the agreement between PNP (curve) and NP+LEMC is better if we use $\lambda_{\mathrm{MSA}}$ for NP+LEMC (filled symbols) than using $\lambda_{\mathrm{D}}$ (open symbols).

\begin{figure}[t]
\begin{center}
\includegraphics*[width=0.49\textwidth]{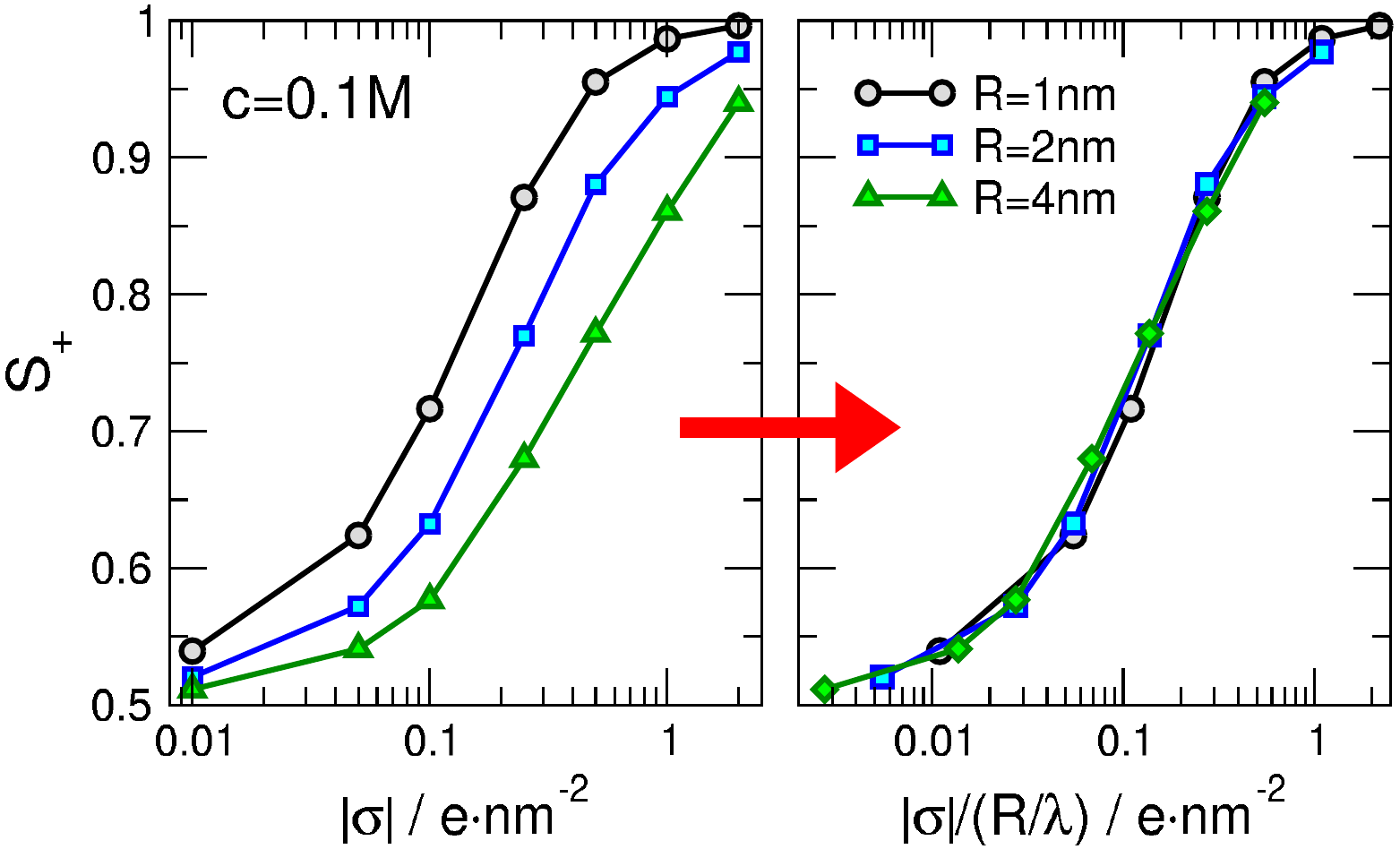}
\end{center}
\caption{Selectivity ($S_{+}$) curves plotted against the surface charge (left panel), and the modified Dukhin number (right panel) for different pore radii (different colors).
The difference between the two panels is that the abscissa is rescaled by dividing $|\sigma|$ with $R/\lambda$.
The unit of $\sigma$ is $e$/nm$^{2}$.
The concentration is constant, $c=0.1$ M.
Only NP+LEMC results are shown with $\lambda=\lambda_{\mathrm{MSA}}$.
}
\label{fig4}
\end{figure}

The radius of the pore was constant in Fig.~\ref{fig2}.
Figure \ref{fig4} shows scaling for varying pore radii at a fixed concentration ($c=0.1$ M).
The left panel shows the $S_{+}$ curves as functions of $|\sigma|$, while the right panel shows them rescaled with $R/\lambda$.
Because $\lambda$ is constant ($c$ is constant) here, in effect this means a rescaling with $R$. 

In summary, scaling works for selectivity using an appropriately chosen scaling parameter that we call the modified Dukhin number, mDu.

%%%%%%%%%%%%%%%%%%%%%%%%%%%%%%%%%%%%%%%%%%%%%%%%%%%%%%%%%%%%%%%%%%%%%%%%%%%%%%%%%%%%%%%%%%%%%%%%%%%%%%%%%%%%%%%%%%%%%%%%%%%%%%%%%%%%%%%%%%%%%%%%%%%%%%%%%%%%%%%%%%%%%%%%%%%%%%%%%%%%%%%%%%%%%%%%%%%%%%%%%%%%%%%%%%%%%%%%%%%%%%%%%%%%%%%%%%%%%%%%%%%%%%%%%%%%%%%%%%%%%%%%%%%%%%%%%%%%%%%%%%%%%%%%%%%%%%%%%%%%%%%%%%%%%%%%%%%%%%%%%

\subsection{Radial selectivity profile to characterize surface and volume conductions }
\label{sec:res:radialsel}

% We can, however, use the capabilities of our modeling study and look at local profiles that are included in the NP equation (Eq.~\ref{eq:np}).

If we have a selectivity value for a given situation of, for example, $S_{+}=0.85$, it does not necessarily mean that there is no volume conduction in the pore.
Similarly, if we have a mildly selective value $S_{+}=0.65$, it does not necessarily mean that there is no surface conduction.
We cannot conclude from the ``global'' selectivity value ($S_{+}$) alone to what degree surface and volume conductions are present in the two respective regions of the pore (close and far from the surface).

How can we characterize the share of the two types of conductions on the basis of the simulation results?
We define the radial selectivity profile as
\begin{equation}
 s_{+}(r) = \dfrac{j_{+}(r)}{j_{+}(r)+j_-(r)},
\end{equation} 
where $j_{+}(r)$ and $j_-(r)$ are the $z$-components of the flux profiles for the cation and the anion, respectively, averaged over the pore in the $z$ dimension (from $-H/2$ to $H/2$).
It is a quantity that depends on $r$ and characterizes to what degree a region of the pore at a distance $r$ from the $z$ axis contributes to the ``global'' selectivity, $S_{+}$.
Note that the average of $s_{+}(r)$ is not equal to $S_{+}$, but we can draw conclusions from the shapes of the curves nevertheless.

\begin{figure*}[t!]
\begin{center}
\includegraphics*[width=0.7\textwidth]{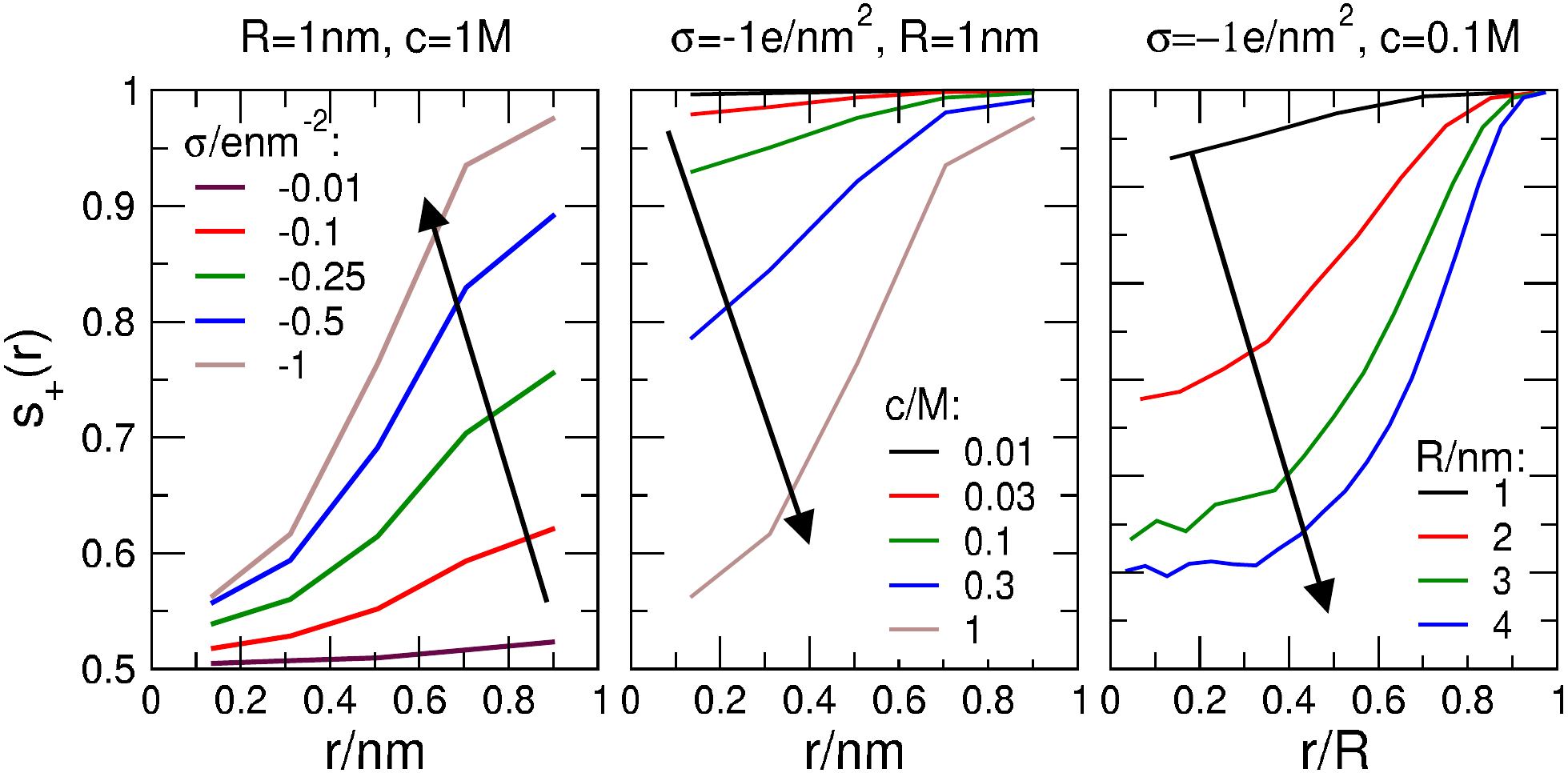}
\end{center}
\caption{The radial selectivity profiles, $s_{+}(r)$, for different surface charges (left panel), concentrations (middle panel), and pore radii (right panel). 
In each case, the other two parameters are kept fixed at the values indicated above the panels.
The arrows show the direction of increasing the respective parameter.
}
\label{fig5}
\end{figure*}

First, let us investigate the effect of the parameters $\sigma$, $c$, and $R$ on the $s_{+}(r)$ function.
Figure \ref{fig5} shows the dependence of these curves on one of these parameters while the other two are kept constant.
The left panel shows that the $s_{+}(r)$ curves are shifted towards larger values as $\sigma$ increases.
A larger surface charge separates cation and anion profiles and excludes anions more efficiently.
The middle panel shows that the $s_{+}(r)$ curves are shifted towards smaller values as $c$ increases.
At larger concentrations, the DLs are thinner, so a wider non-selective bulk region forms in the middle of the pore thus bringing selectivity down.
The right panel shows that the $s_{+}(r)$ profiles are shifted towards smaller values as $R$ increases.
In wider pores there is more space for the formation of a bulk region in the middle of the pore.

Figure \ref{fig5} also shows that there are cases when both a surface conduction region (large selectivity near the surface) and a volume conduction region (small selectivity in the center of the pore) are present.
For moderate overlap (left panel, $R/\lambda=0.912$) it occurs if $\sigma $ is large enough because large $\sigma $ increases selectivity near the wall.
For a relatively large surface charge (middle panel, $\sigma=-1$ $e$/nm$^{2}$), it occurs at large concentrations that produces small screening lengths and confine the DL near the wall region.
The pore radius has similar effects with the difference that now we tune the $R/\lambda$ ratio with $R$ (right panel).

\begin{figure*}[t!]
\begin{center}
\textsf{(A)}\includegraphics*[width=0.7\textwidth]{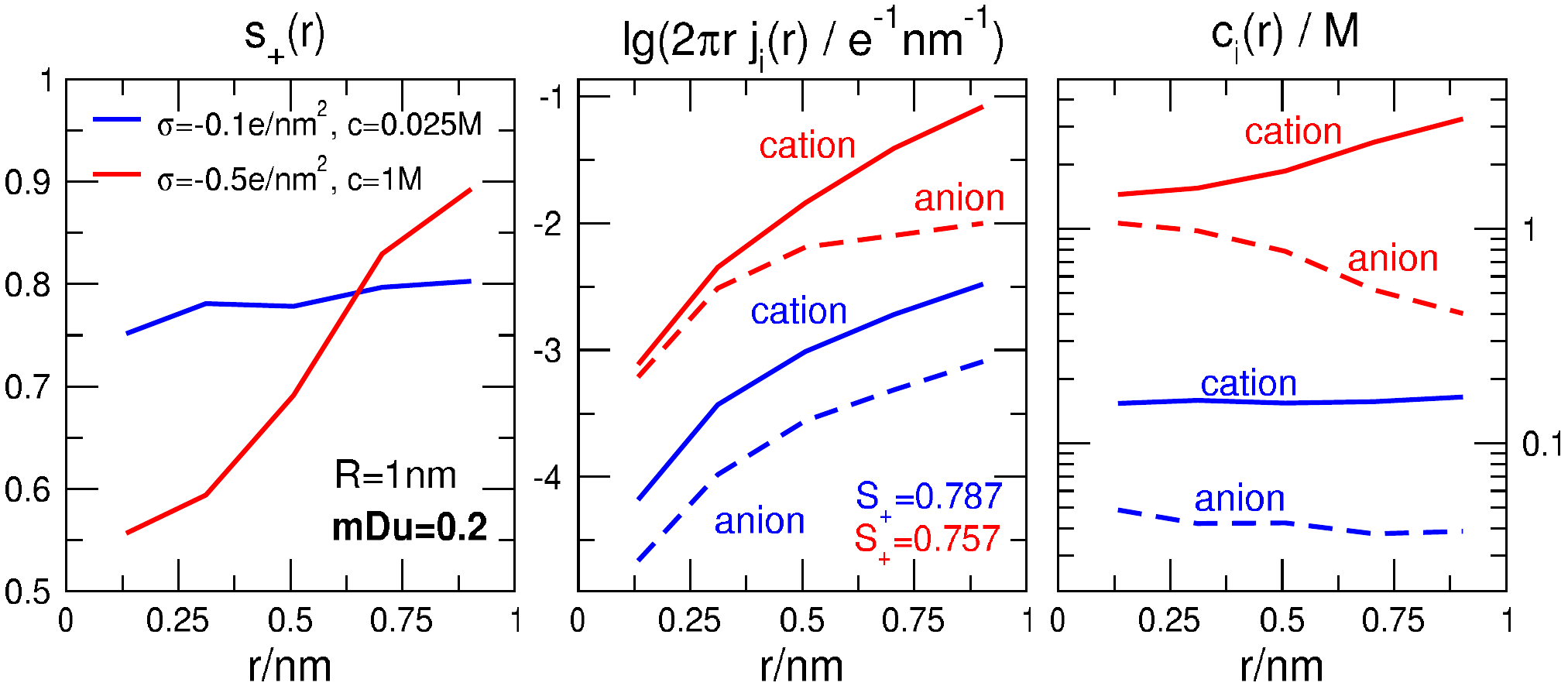}
\end{center}
% \caption{
% }
% \label{fig5}
% \end{figure*} 
% 
% \begin{figure*}[t!]
\begin{center}
\textsf{(B)}\includegraphics*[width=0.7\textwidth]{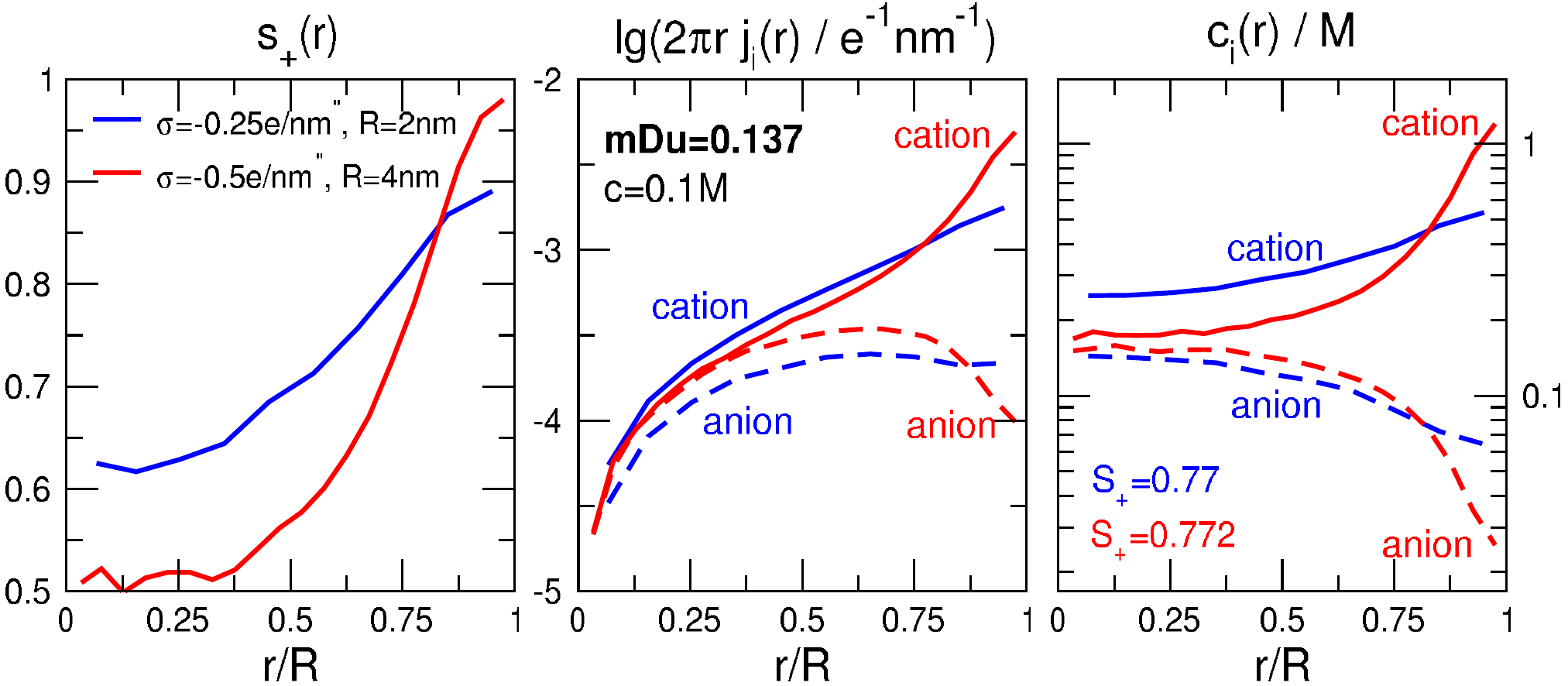}
\end{center}
\caption{Radial selectivity profiles ($s_{+}(r)$, left panel), radial flux profiles (the $z$-coordinate averaged over the pore in the axial direction) multiplied by $2\pi r$ ($2\pi r j_{i}(r)$, middle panel), and radial concentration profiles ($c_{i}(r)$, right panel) for parameters indicated in the figure.
(A) The pore radius is fixed ($R=1$ nm), while $\sigma$ and $c$ are changed in a way that mDu is the same ($\mathrm{mDu}=0.2$ nm$^{-2}$) for the two sets of curves (red and blue).
(B) The concentration is fixed ($c=0.1$ M), while $\sigma$ and $R$ are changed in a way that mDu is the same ($\mathrm{mDu}=0.137$ nm$^{-2}$) for the two sets of curves (red and blue).
In the middle and right panels, solid and dashed lines refer to cations and anions, respectively.
 }
\label{fig6}
\end{figure*}

Figure~\ref{fig6} illustrates the notion that a given ``global'' selectivity can be achieved with different sets of parameters that correspond to different $s_{+}(r)$ profiles.
The figure shows curves for the same mDu parameter but with different combinations of $\sigma$, $R$, and $c$ producing the same mDu.

Figure \ref{fig6}A shows results for two simulations that, due to scaling, produce about the same ``global'' selectivity, $S_{+}=0.787$ (blue) and $0.757$ (red). 
Both the blue and red curves have the same mDu number ($\mathrm{mDu}=0.2$ nm$^{-2}$) and the same pore radius ($R=1$ nm).
The blue curve is for a smaller surface charge ($\sigma =-0.1$ $e$/nm$^{2}$) and a smaller concentration ($c=0.025$ M), while the red curve is for a larger surface charge ($\sigma =-0.5$ $e$/nm$^{2}$) and a larger concentration ($c=1$ M).

The $s_{+}(r)$ curves (left panel) show that approximately the  same $S_{+}$ value can be achieved in two ways. 
In the case of the blue curve, the DL overlaps (small $c$, large $\lambda$) so the coion is excluded over the whole cross section and selectivity is at a high level even if $\sigma$ is relatively small.
In the case of the red curve, the DL overlaps less (large $c$, small $\lambda$) so a less selective bulk region is formed in the middle, but the larger surface charge ($\sigma=-0.5$ $e$/nm$^{2}$) produces a large counterion-coion separation near the wall and pulls up selectivity there.

The concentration and flux profiles in the other two panels show how this comes about.
Different degree of overlapping can be seen better in the concentration profiles (right panel).
In the middle panel, the $2\pi r j_{i}(r)$ curves are shown on a logarithmic scale.
% Plotting this way, the area under the curves corresponds to the total current ($I_{i}$), while 
The gap between the cation (solid) and anion (dashed) profiles shows the degree of selectivity at $r$.
In the case of the blue curves, the gap is the same all along the pore in the radial dimension, while in the case of the red curves, the gap opens as the pore wall is approached.

Figure \ref{fig6}B shows results for two simulations that have the same mDu number ($\mathrm{mDu}=0.137$ nm$^{-2}$) and the same concentration ($c=0.1$ M).
Because $R$ changes, we show the profiles as functions of $r/R$ here.
They produce approximately the same selectivities, $S_{+}=0.77$ (blue) and $0.772$ (red).
The blue curve is for a smaller surface charge ($\sigma =-0.25$ $e$/nm$^{2}$) and a smaller pore radius ($R=2$ nm), while the red curve is for a larger surface charge ($\sigma =-0.5$ $e$/nm$^{2}$) and a larger pore radius ($R=4$ nm).

The $s_{+}(r)$ curves (left panel) show similar behavior as in Fig.~\ref{fig6}A.
In the case of the blue curve, the DL overlaps (small $R$) so there is less space for the formation of a bulk region in the middle.
Therefore, the selectivity is uniformly large even if $\sigma$ is relatively small.
In the case of the red curve, the degree of overlap of the DL is smaller (large $R$) so a less selective bulk region is formed in the middle, but the larger surface charge ($\sigma=-0.5$ $e$/nm$^{2}$) produces a large counterion-coion separation near the wall and pulls up selectivity there.

The concentration profiles (right panel) show the different degrees of overlap in the two cases.
The flux profiles (middle panel) show as the gap opens more widely in the case of the red curves than in the case of the blue curves.

%%%%%%%%%%%%%%%%%%%%%%%%%%%%%%%%%%%%%%%%%%%%%%%%%%%%%%%%%%%%%%%%%%%%%%%%%%%%%%%%%%%%%%%%%%%%%%%%%%%%%%%%%%%%%%%%%%%%%%%%%%%%%%%%%%%%%%%%%%%%%%%%%%%%%%%%%%%%%%%%%%%%%%%%%%%%%%%%%%%%%%%%%%%%%%%%%%%%%%%%%%%%%%%%%%%%%%%%%%%%%%%%%%%%%%%%%%%%%%%%%%%%%%%%%%%%%%%%%%%%%%%%%%%%%%%%%%%%%%%%%%%%%%%%%%%%%%%%%%%%%%%%%%%%%%%%%%%%%%%%%

\subsection{Transition point between surface and volume conductions}
\label{sec:inflexion}

By plotting selectivity as a function of mDu using a logarithmic scale on the mDu axis, we get a curve with an inflection point that belongs to a selectivity $S_{+}\approx 0.75$.
If we fit a sigmoid curve too all the points of Fig.~\ref{fig2}, the resulting selectivity value for the inflection point is $S_{+}\approx 0.753$.
The inflection point offers itself as a transition point separating ``rather non-selective'' and ``rather selective'' regions.
The  derivative of the $S_{+}$ curve as a function of $\lg(\mathrm{mDu})$ is maximal in the inflection point where small changes in mDu lead to relatively large changes in selectivity.

\begin{figure*}[t!]
\begin{center}
\includegraphics*[width=0.7\textwidth]{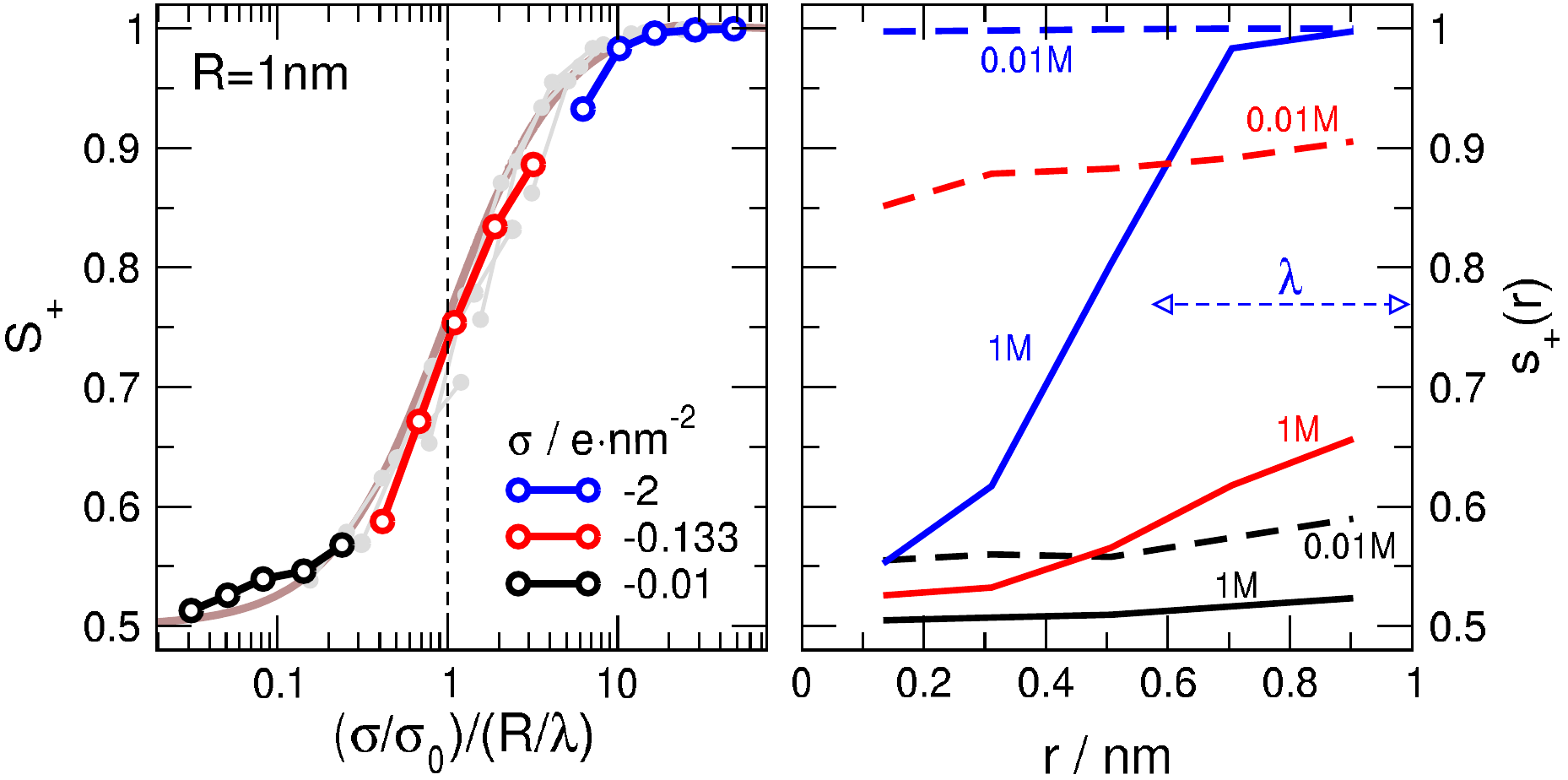}
\end{center}
\caption{
Selectivity, $S_{+}$, plotted against the modified Dukhin number normalized as , $(\sigma /\sigma_{0})/e(R/ \lambda )$, where $\sigma_{0}\equiv -0.133$ $e$/nm$^{2}$ (left panel). 
Different colors mean different surface charges, while points with a given color correspond to different concentrations.
The values $\sigma=-0.01$, $-0.133$, and $-2$ $e$/nm$^{2}$ are highlighted with different colors (black, red, and blue), while other surface charges are shown with gray for reference.
The right panel shows the radial selectivity profiles, $s_{+}(r)$, for these surface charges for concentrations $c=0.01$ and $1$ M (filled and open symbols).
The MSA screening length for $c=1$ M is indicated by the horizontal dashed line with arrows.
}
\label{fig7}
\end{figure*}

The inflection point is at $0.133$ nm$^{-2}$, so if we normalize $\sigma$ in mDu with $\sigma_{0}\equiv-0.133$ $e$/nm$^{2}$ as $(\sigma/ \sigma_{0}) / (R/\lambda)$, we obtain a rescaled mDu that has the inflection point at $1$ as shown by the left panel of Fig.~\ref{fig7}.
This panel shows the same data that the right panel of Fig.~\ref{fig2} showed with mDu rescaled and visualizing the results differently.
A given color now corresponds to a given surface charge with concentration changing in the range $c=0.01-1$ M.
Black, red, and blue colors correspond to surface charges $\sigma=-0.01$, $-0.133$, and $-2$ $e$/nm$^{2}$, respectively.

The slope is maximal in the case of $\sigma=-0.133$ $e$/nm$^{2}$ (red curve).
$S_{+}$ changes in a wide range with varying $R/\lambda$ with $S_{+}=0.75$ being the mean value.
At this surface charge, $S_{+}$ is far from the limiting values $0.5$ and $1$, at least for the simulated values of $c$ and $R$.
Low and large surface charges ($\sigma=-0.01$ and $-2$ $e$/nm$^{2}$, black and blue curves), on the other hand, are restricted to small ($\sim 0.5$) and large ($\sim 1$) selectivities.

The right panel of Fig.~\ref{fig7} shows the $s_{+}(r)$ selectivity profiles for the largest ($c=1$ M, dashed lines) and smallest ($c=0.01$ M, solid lines) simulated concentrations for the three surface charges in focus.
Selectivity is small for $\sigma=-0.01$ $e$/nm$^{2}$ (black) even for $c=0.01$ M, where DL overlap tend to increase the selectivity.
For the value $\sigma=-0.133$ $e$/nm$^{2}$, selectivity changes in a wide range as a result of changing the concentration between $c=0.01$ and $1$ M.
In the case of $\sigma=-2$ $e$/nm$^{2}$, the large DL overlap at $c=0.01$ M results in a closely perfect selectivity.

The curve for $\sigma=-2$ $e$/nm$^{2}$ and $c=1$ M is more interesting.
A clean separation of volume and surface conductions can be observed in this case because the DL is restricted to the surface due to the small value of $\lambda$ (large $c$).
This curve corresponds to a ``global'' selectivity $S_{+}=0.933$ that is not readily apparent from the $s_{+}(r)$ curve.
As noted earlier, $S_{+}$ is not the average of $s_{+}(r)$; regions near the wall contribute more to current, and, thus, to $S_{+}$, because those regions have larger cross section (see middle panels of Fig.~\ref{fig6}).
This result emphasizes that surface conduction tends to have a larger weight compared to the volume conduction due to the fact that cross sections are larger near the pore wall than near the centerline.
Also, these results imply that a sufficiently large surface charge is needed if we want to obtain a selective surface-conduction region near the wall.

\section{Summary}

We showed that cation selectivity as our chosen device function scales properly with the scaling parameter inspired by the Dukhin number.
Continuing the tradition of saluting Dukhin, we call this paremeter the modified Dukhin number. 
This new parameter differs from the original one that it contains the $R/\lambda$ ratio instead of the $Rc$ product. 
As a consequence, it is proportional to $c^{-1/2}$ instead of $c^{-1}$.

The modified version satisfies scaling and characterizes selectivity in the intermediate regime, where both surface and bulk conductions may be present and the pore may be neither perfectly selective, nor perfectly non-selective.
Our modeling study using the LEMC simulation method and the PNP theory  provides the radial flux profiles from which the radial selectivity profile can be computed.
These profiles show in which region of the nanopore the surface or the volume conduction dominates for a given combination of the variables $\sigma$, $R$, and $c$.

To characterize the transition point that separates the ``rather selective'' and the ``rather non-selective'' state points, the inflection point of the $S_{+}$ vs.\ $\lg(\mathrm{mDu})$ function is a natural, and mathematically well defined, choice.
The inflection point depends on system parameters ($d_{i}$, $H$, $U$, $T$, and $\epsilon$), but otherwise it is a well-defined universal feature of this system for 1:1 electrolytes.  

For electrolytes containing multivalent ions (2:2, 2:1, and 3:1, for example) interesting phenomena beyond the mean-field treatment may occur due to strong ionic correlations such as overcharging and charge inversion.~\cite{fertig_pccp_2020}
These will be reported in subsequent publications.

\section*{Acknowledgements}
\label{sec:ack}

We gratefully acknowledge  the financial support of the National Research, Development and Innovation Office -- NKFIH K124353. 
Present article was published in the frame of the project GINOP-2.3.2-15-2016-00053 (``Development of engine fuels with high hydrogen content in their molecular structures (contribution to sustainable mobility)'').

%%%%%%%%%%%%%%%%%%%%%%%%%%%%%%%%%%%%%%%%%%%%%%%%%%%%%%%%%%%%%%%%%%%%%%%%%%%%%%%%%%%%%%%%%%%%%%%%%%%%%%%%%%%%%%%%%%%%%%%%%%%%%%%%%%%%%%%%%%%%%%%%%%%%%%%%%%%%%%%%%%%%%%%%%%%%%%%%%%%%%%%%%%%%%%%%%%%%%%%%%%%%%%%%%%%%%%%%%%%%%%%%%%%%%%%%%%%%%%%%%%%%%%%%%%%%%%%%%%%%%%%%%%%%%%%%%%%%%%%%%%%%%%%%%%%%%%%%%%%%%%%%%%%%%%%%%%%%%%%%%

\appendix
\section{Computational methods}
\label{sec:me4hods}

In this work, we use two methods that have the common denominator that both apply the Nernst-Planck (NP) transport equation~\cite{nernst_zpc_1888,planck_apc_1890} to compute the ionic flux:
\begin{equation}
 \mathbf{j}_{i}(\mathbf{r})= -\frac{1}{kT} D_{i}(\mathbf{r})c_{i}(\mathbf{r})\nabla \mu_{i}(\mathbf{r}),
 \label{eq:np}
\end{equation} 
where $\mathbf{j}_{i}(\mathbf{r})$, $D_{i}(\mathbf{r})$, $c_{i}(\mathbf{r})$, and $\mu_{i}(\mathbf{r})$ are the flux density, the diffusion coefficent profile, the concentration profile, and the electrochemical potential profile of ionic species $i$, respectively.
To make use of this equation, we need a relation between the concentration profile, $c_{i}(\mathbf{r})$, and the electrochemical potential profile, $\mu_{i}(\mathbf{r})$.

The LEMC method~\cite{boda_jctc_2012} is a particle simulation technique devised for a non-equilibrium situation, where $\mu_{i}(\mathbf{r})$ is not constant globally.
We divide the simulation into small volume elements, $V^{\alpha}$, and assume local thermodynamic equilibrium in each, namely, we assign $\mu_{i}^{\alpha}$ values to each.
Then, we perform particle displacements and particle insertions/deletions with the same acceptannce criteria as we do in a GCMC simulation, but with the volume, particle number ($N_{i}^{\alpha}$), and chemical potential of the volume element in which we peform an MC step.
Therefore, the LEMC technique is an adaptation of the GCMC technique for a system that is not at equilibrium globally.

The resulting method, coined NP+LEMC, solves the problem iteratively on the basis of the scheme
\begin{equation}
\mu^{\alpha}_{i}[n] \, \xrightarrow{\mathrm{LEMC}}  \,  c^{\alpha}_{i}[n] \,  \xrightarrow{\mathrm{NP}} \,  \mathbf{j}^{\alpha}_{i}[n] \, 
\xrightarrow{\nabla \cdot \mathbf{j}=0} 
\,\, \mu^{\alpha}_{i}[n+1] ,
\label{eq:circle}
\end{equation} 
where $c_{i}^{\alpha}[n]$ is the concentration in volume element $V^{\alpha}$ obtained from an LEMC simulation in the $n$th iteration.
The chemical potential for the next $[n{+}1]$th iteration is obtained by assuming that the flux $\mathbf{j}_{i}^{\alpha}$ computed from it and from $c_{i}^{\alpha}[n]$ satisfies the continuity equation, $\nabla \cdot \mathbf{j}_{i}=0$.
Details are found in Refs.~\onlinecite{boda_jctc_2012,boda_jml_2014,boda_arcc_2014,fertig_hjic_2017}.

The importance of the LEMC technique is that it can take into account the correlations between ions beyond the mean-field approximation including the finite size of ions.

In the Poisson-Nernst-Planck (PNP) theory, we relate $c_{i}(\mathbf{r})$ to $\mu_{i}(\mathbf{r})$ via the Poisson-Boltzmann (PB) theory where the ions are modeled as point charges interacting with the average electrical potential, $\Phi(\mathbf{r})$,  exerted by all the ions in the system.
In this mean-field approach, the electrolyte is assumed to be an ideal solution with the electrochemical potential
\begin{equation*}
 \mu_{i}(\mathbf{r}) = \mu_{i}^{0} + kT\ln c_{i}(\mathbf{r}) + z_{i}e\Phi(\mathbf{r}),
\end{equation*}
where $\mu_{i}^{0}$ is a reference chemical potential independent of the location.
Note that an excess chemical potential $\mu_{i}^{\mathrm{ex}}(\mathbf{r})$, is added to this expression when ionic correlations are taken into account as they are in the LEMC method.
Poisson's equation and the continuity equation are also satisfied in the solution of the PNP theory.

Here, we solve the system with the Scharfetter–Gummel scheme.~\cite{gummel1964self} 
A 2D finite element method is used with $20-60$ thousands elements in a triangular mesh.
Details are found in Ref.~\onlinecite{matejczyk_jcp_2017}.

The constant $\sigma$ surface charge is assured via a Neumann boundary condition in PNP, while fractional point charges are placed on a rectangular grid of width $0.2$ nm in LEMC.

In both methods, proper boundary conditions were applied in the two baths at the wall of the cylinder that confines the finite simulation cell. 
Dirichlet boundary conditions were applied for $\Phi(\mathrm{r})$; the difference of the applied potential on the two sides of the membrane specifies the applied voltage, $U$.
Bath concentrations were assumed to be the same on the two sides of the membrane, $c$, though the methodology would be able to handle asymmetrical systems as well. 
In the 1:1 electrolyte considered here $c_{+}=c_-=c$ in both baths.

For the diffusion coefficient profile, $D_{i}(\mathbf{r})$, we use a piecewise constant function, where the value in the baths is $1.334\times 10^{-9}$ m$^{2}$s$^{-1}$ for both ionic species, while it is the tenth of that inside the pore, $D_{i}^{\mathrm{pore}}$, as in our earlier works.~\cite{matejczyk_jcp_2017,madai_jcp_2017,madai_pccp_2018,fertig_jpcc_2019,fertig_pccp_2020}
These particular choices do not qualitatively affect our conclusions.

In this model calculation, where $d_{+}=d_-$ and $D_{+}(\mathbf{r})=D_-(\mathbf{r})$ for a 1:1 electrolyte, so $S_{+}=0.5$ exactly for a perfectly non-selective pore ($\sigma=0$, for example)

% \bibliography{bib/nanopore,bib/book,bib/own}

\begin{thebibliography}{32}%
\makeatletter
\providecommand \@ifxundefined [1]{%
 \@ifx{#1\undefined}
}%
\providecommand \@ifnum [1]{%
 \ifnum #1\expandafter \@firstoftwo
 \else \expandafter \@secondoftwo
 \fi
}%
\providecommand \@ifx [1]{%
 \ifx #1\expandafter \@firstoftwo
 \else \expandafter \@secondoftwo
 \fi
}%
\providecommand \natexlab [1]{#1}%
\providecommand \enquote  [1]{``#1''}%
\providecommand \bibnamefont  [1]{#1}%
\providecommand \bibfnamefont [1]{#1}%
\providecommand \citenamefont [1]{#1}%
\providecommand \href@noop [0]{\@secondoftwo}%
\providecommand \href [0]{\begingroup \@sanitize@url \@href}%
\providecommand \@href[1]{\@@startlink{#1}\@@href}%
\providecommand \@@href[1]{\endgroup#1\@@endlink}%
\providecommand \@sanitize@url [0]{\catcode `\\12\catcode `\$12\catcode
  `\&12\catcode `\#12\catcode `\^12\catcode `\_12\catcode `\%12\relax}%
\providecommand \@@startlink[1]{}%
\providecommand \@@endlink[0]{}%
\providecommand \url  [0]{\begingroup\@sanitize@url \@url }%
\providecommand \@url [1]{\endgroup\@href {#1}{\urlprefix }}%
\providecommand \urlprefix  [0]{URL }%
\providecommand \Eprint [0]{\href }%
\providecommand \doibase [0]{http://dx.doi.org/}%
\providecommand \selectlanguage [0]{\@gobble}%
\providecommand \bibinfo  [0]{\@secondoftwo}%
\providecommand \bibfield  [0]{\@secondoftwo}%
\providecommand \translation [1]{[#1]}%
\providecommand \BibitemOpen [0]{}%
\providecommand \bibitemStop [0]{}%
\providecommand \bibitemNoStop [0]{.\EOS\space}%
\providecommand \EOS [0]{\spacefactor3000\relax}%
\providecommand \BibitemShut  [1]{\csname bibitem#1\endcsname}%
\let\auto@bib@innerbib\@empty
%</preamble>
\bibitem [{\citenamefont {Fertig}\ \emph {et~al.}(2019)\citenamefont {Fertig},
  \citenamefont {Matejczyk}, \citenamefont {Valisk{\'{o}}}, \citenamefont
  {Gillespie},\ and\ \citenamefont {Boda}}]{fertig_jpcc_2019}%
  \BibitemOpen
  \bibfield  {author} {\bibinfo {author} {\bibfnamefont {D.}~\bibnamefont
  {Fertig}}, \bibinfo {author} {\bibfnamefont {B.}~\bibnamefont {Matejczyk}},
  \bibinfo {author} {\bibfnamefont {M.}~\bibnamefont {Valisk{\'{o}}}}, \bibinfo
  {author} {\bibfnamefont {D.}~\bibnamefont {Gillespie}}, \ and\ \bibinfo
  {author} {\bibfnamefont {D.}~\bibnamefont {Boda}},\ }\href {\doibase
  10.1021/acs.jpcc.9b07574} {\bibfield  {journal} {\bibinfo  {journal} {J.
  Phys. Chem. C}\ }\textbf {\bibinfo {volume} {123}},\ \bibinfo {pages} {28985}
  (\bibinfo {year} {2019})}\BibitemShut {NoStop}%
\bibitem [{\citenamefont {Blum}(1975)}]{blum_mp_1975}%
  \BibitemOpen
  \bibfield  {author} {\bibinfo {author} {\bibfnamefont {L.}~\bibnamefont
  {Blum}},\ }\href {\doibase 10.1080/00268977500103051} {\bibfield  {journal}
  {\bibinfo  {journal} {Mol. Phys.}\ }\textbf {\bibinfo {volume} {30}},\
  \bibinfo {pages} {1529} (\bibinfo {year} {1975})}\BibitemShut {NoStop}%
\bibitem [{\citenamefont {Blum}\ and\ \citenamefont
  {Hoeye}(1977)}]{blum_jcp_1977}%
  \BibitemOpen
  \bibfield  {author} {\bibinfo {author} {\bibfnamefont {L.}~\bibnamefont
  {Blum}}\ and\ \bibinfo {author} {\bibfnamefont {J.~S.}\ \bibnamefont
  {Hoeye}},\ }\href {\doibase 10.1021/j100528a019} {\bibfield  {journal}
  {\bibinfo  {journal} {J. Phys. Chem.}\ }\textbf {\bibinfo {volume} {81}},\
  \bibinfo {pages} {1311} (\bibinfo {year} {1977})}\BibitemShut {NoStop}%
\bibitem [{\citenamefont {Nonner}, \citenamefont {Catacuzzeno},\ and\
  \citenamefont {Eisenberg}(2000)}]{nonner_bj_2000}%
  \BibitemOpen
  \bibfield  {author} {\bibinfo {author} {\bibfnamefont {W.}~\bibnamefont
  {Nonner}}, \bibinfo {author} {\bibfnamefont {L.}~\bibnamefont {Catacuzzeno}},
  \ and\ \bibinfo {author} {\bibfnamefont {B.}~\bibnamefont {Eisenberg}},\
  }\href {\doibase 10.1016/s0006-3495(00)76446-0} {\bibfield  {journal}
  {\bibinfo  {journal} {Biophys. J.}\ }\textbf {\bibinfo {volume} {79}},\
  \bibinfo {pages} {1976} (\bibinfo {year} {2000})}\BibitemShut {NoStop}%
\bibitem [{\citenamefont {M\'adai}\ \emph {et~al.}(2018)\citenamefont
  {M\'adai}, \citenamefont {Matejczyk}, \citenamefont {Dallos}, \citenamefont
  {Valisk\'o},\ and\ \citenamefont {Boda}}]{madai_pccp_2018}%
  \BibitemOpen
  \bibfield  {author} {\bibinfo {author} {\bibfnamefont {E.}~\bibnamefont
  {M\'adai}}, \bibinfo {author} {\bibfnamefont {B.}~\bibnamefont {Matejczyk}},
  \bibinfo {author} {\bibfnamefont {A.}~\bibnamefont {Dallos}}, \bibinfo
  {author} {\bibfnamefont {M.}~\bibnamefont {Valisk\'o}}, \ and\ \bibinfo
  {author} {\bibfnamefont {D.}~\bibnamefont {Boda}},\ }\href {\doibase
  10.1039/c8cp03918f} {\bibfield  {journal} {\bibinfo  {journal} {Phys. Chem.
  Chem. Phys.}\ }\textbf {\bibinfo {volume} {20}},\ \bibinfo {pages} {24156}
  (\bibinfo {year} {2018})}\BibitemShut {NoStop}%
\bibitem [{\citenamefont {Bazant}, \citenamefont {Thornton},\ and\
  \citenamefont {Ajdari}(2004)}]{bazant_pre_2004}%
  \BibitemOpen
  \bibfield  {author} {\bibinfo {author} {\bibfnamefont {M.~Z.}\ \bibnamefont
  {Bazant}}, \bibinfo {author} {\bibfnamefont {K.}~\bibnamefont {Thornton}}, \
  and\ \bibinfo {author} {\bibfnamefont {A.}~\bibnamefont {Ajdari}},\ }\href
  {\doibase 10.1103/physreve.70.021506} {\bibfield  {journal} {\bibinfo
  {journal} {Phys. Rev. E}\ }\textbf {\bibinfo {volume} {70}},\ \bibinfo
  {pages} {021506} (\bibinfo {year} {2004})}\BibitemShut {NoStop}%
\bibitem [{\citenamefont {Chu}\ and\ \citenamefont
  {Bazant}(2006)}]{chu_pre_2006}%
  \BibitemOpen
  \bibfield  {author} {\bibinfo {author} {\bibfnamefont {K.~T.}\ \bibnamefont
  {Chu}}\ and\ \bibinfo {author} {\bibfnamefont {M.~Z.}\ \bibnamefont
  {Bazant}},\ }\href {\doibase 10.1103/physreve.74.011501} {\bibfield
  {journal} {\bibinfo  {journal} {Phys. Rev. E}\ }\textbf {\bibinfo {volume}
  {74}},\ \bibinfo {pages} {011501} (\bibinfo {year} {2006})}\BibitemShut
  {NoStop}%
\bibitem [{\citenamefont {Bocquet}\ and\ \citenamefont
  {Charlaix}(2010)}]{bocquet_chemsocrev_2010}%
  \BibitemOpen
  \bibfield  {author} {\bibinfo {author} {\bibfnamefont {L.}~\bibnamefont
  {Bocquet}}\ and\ \bibinfo {author} {\bibfnamefont {E.}~\bibnamefont
  {Charlaix}},\ }\href {\doibase 10.1039/b909366b} {\bibfield  {journal}
  {\bibinfo  {journal} {Chem. Soc. Rev.}\ }\textbf {\bibinfo {volume} {39}},\
  \bibinfo {pages} {1073} (\bibinfo {year} {2010})}\BibitemShut {NoStop}%
\bibitem [{\citenamefont {Bikerman}(1940)}]{bikerman_1940}%
  \BibitemOpen
  \bibfield  {author} {\bibinfo {author} {\bibfnamefont {J.~J.}\ \bibnamefont
  {Bikerman}},\ }\href {\doibase 10.1039/tf9403500154} {\bibfield  {journal}
  {\bibinfo  {journal} {Trans. Farad. Soc.}\ }\textbf {\bibinfo {volume}
  {35}},\ \bibinfo {pages} {154} (\bibinfo {year} {1940})}\BibitemShut
  {NoStop}%
\bibitem [{\citenamefont {Dukhin}(1993)}]{dukhin_advcollsci_1993}%
  \BibitemOpen
  \bibfield  {author} {\bibinfo {author} {\bibfnamefont {S.}~\bibnamefont
  {Dukhin}},\ }\href {\doibase 10.1016/0001-8686(93)80021-3} {\bibfield
  {journal} {\bibinfo  {journal} {Adv. Coll. Interf. Sci.}\ }\textbf {\bibinfo
  {volume} {44}},\ \bibinfo {pages} {1} (\bibinfo {year} {1993})}\BibitemShut
  {NoStop}%
\bibitem [{\citenamefont {Lyklema}\ \emph {et~al.}(1995)\citenamefont
  {Lyklema}, \citenamefont {de~Keizer}, \citenamefont {Bijsterbosch},
  \citenamefont {Fleer},\ and\ \citenamefont {(Eds.)}}]{lyklema_book_1995}%
  \BibitemOpen
  \bibfield  {author} {\bibinfo {author} {\bibfnamefont {J.~J.}\ \bibnamefont
  {Lyklema}}, \bibinfo {author} {\bibfnamefont {A.}~\bibnamefont {de~Keizer}},
  \bibinfo {author} {\bibfnamefont {B.}~\bibnamefont {Bijsterbosch}}, \bibinfo
  {author} {\bibfnamefont {G.}~\bibnamefont {Fleer}}, \ and\ \bibinfo {author}
  {\bibfnamefont {M.~C.~S.}\ \bibnamefont {(Eds.)}},\ }\href
  {http://gen.lib.rus.ec/book/index.php?md5=145e94c05719008b3b7f73e2a5da6e89}
  {\emph {\bibinfo {title} {Solid-Liquid Interfaces}}},\ Fundamentals of
  Interface and Colloid Science 2\ (\bibinfo  {publisher} {Elsevier, Academic
  Press},\ \bibinfo {year} {1995})\BibitemShut {NoStop}%
\bibitem [{\citenamefont {Khair}\ and\ \citenamefont
  {Squires}(2008)}]{khair_jfm_2008}%
  \BibitemOpen
  \bibfield  {author} {\bibinfo {author} {\bibfnamefont {A.~S.}\ \bibnamefont
  {Khair}}\ and\ \bibinfo {author} {\bibfnamefont {T.~M.}\ \bibnamefont
  {Squires}},\ }\href {\doibase 10.1017/s002211200800390x} {\bibfield
  {journal} {\bibinfo  {journal} {J. Fluid Mech.}\ }\textbf {\bibinfo {volume}
  {615}},\ \bibinfo {pages} {323} (\bibinfo {year} {2008})}\BibitemShut
  {NoStop}%
\bibitem [{\citenamefont {Das}\ and\ \citenamefont
  {Chakraborty}(2010)}]{das_langmuir_2010}%
  \BibitemOpen
  \bibfield  {author} {\bibinfo {author} {\bibfnamefont {S.}~\bibnamefont
  {Das}}\ and\ \bibinfo {author} {\bibfnamefont {S.}~\bibnamefont
  {Chakraborty}},\ }\href {\doibase 10.1021/la1009237} {\bibfield  {journal}
  {\bibinfo  {journal} {Langmuir}\ }\textbf {\bibinfo {volume} {26}},\ \bibinfo
  {pages} {11589} (\bibinfo {year} {2010})}\BibitemShut {NoStop}%
\bibitem [{\citenamefont {Zangle}, \citenamefont {Mani},\ and\ \citenamefont
  {Santiago}(2010)}]{zangle_csr_2010}%
  \BibitemOpen
  \bibfield  {author} {\bibinfo {author} {\bibfnamefont {T.~A.}\ \bibnamefont
  {Zangle}}, \bibinfo {author} {\bibfnamefont {A.}~\bibnamefont {Mani}}, \ and\
  \bibinfo {author} {\bibfnamefont {J.~G.}\ \bibnamefont {Santiago}},\ }\href
  {\doibase 10.1039/b902074h} {\bibfield  {journal} {\bibinfo  {journal} {Chem.
  Soc. Rev.}\ }\textbf {\bibinfo {volume} {39}},\ \bibinfo {pages} {1014}
  (\bibinfo {year} {2010})}\BibitemShut {NoStop}%
\bibitem [{\citenamefont {Lee}\ \emph {et~al.}(2012)\citenamefont {Lee},
  \citenamefont {Joly}, \citenamefont {Siria}, \citenamefont {Biance},
  \citenamefont {Fulcrand},\ and\ \citenamefont {Bocquet}}]{lee_nanolett_2012}%
  \BibitemOpen
  \bibfield  {author} {\bibinfo {author} {\bibfnamefont {C.}~\bibnamefont
  {Lee}}, \bibinfo {author} {\bibfnamefont {L.}~\bibnamefont {Joly}}, \bibinfo
  {author} {\bibfnamefont {A.}~\bibnamefont {Siria}}, \bibinfo {author}
  {\bibfnamefont {A.-L.}\ \bibnamefont {Biance}}, \bibinfo {author}
  {\bibfnamefont {R.}~\bibnamefont {Fulcrand}}, \ and\ \bibinfo {author}
  {\bibfnamefont {L.}~\bibnamefont {Bocquet}},\ }\href {\doibase
  10.1021/nl301412b} {\bibfield  {journal} {\bibinfo  {journal} {Nano Lett.}\
  }\textbf {\bibinfo {volume} {12}},\ \bibinfo {pages} {4037} (\bibinfo {year}
  {2012})}\BibitemShut {NoStop}%
\bibitem [{\citenamefont {Yeh}\ \emph {et~al.}(2014)\citenamefont {Yeh},
  \citenamefont {Wang}, \citenamefont {Chang},\ and\ \citenamefont
  {Yang}}]{yeh_ijc_2014}%
  \BibitemOpen
  \bibfield  {author} {\bibinfo {author} {\bibfnamefont {H.-C.}\ \bibnamefont
  {Yeh}}, \bibinfo {author} {\bibfnamefont {M.}~\bibnamefont {Wang}}, \bibinfo
  {author} {\bibfnamefont {C.-C.}\ \bibnamefont {Chang}}, \ and\ \bibinfo
  {author} {\bibfnamefont {R.-J.}\ \bibnamefont {Yang}},\ }\href {\doibase
  10.1002/ijch.201400079} {\bibfield  {journal} {\bibinfo  {journal} {Israel J.
  Chem.}\ }\textbf {\bibinfo {volume} {54}},\ \bibinfo {pages} {1533} (\bibinfo
  {year} {2014})}\BibitemShut {NoStop}%
\bibitem [{\citenamefont {Ma}\ \emph {et~al.}(2017)\citenamefont {Ma},
  \citenamefont {Guo}, \citenamefont {Jia},\ and\ \citenamefont
  {Xie}}]{ma_acssens_2017}%
  \BibitemOpen
  \bibfield  {author} {\bibinfo {author} {\bibfnamefont {Y.}~\bibnamefont
  {Ma}}, \bibinfo {author} {\bibfnamefont {J.}~\bibnamefont {Guo}}, \bibinfo
  {author} {\bibfnamefont {L.}~\bibnamefont {Jia}}, \ and\ \bibinfo {author}
  {\bibfnamefont {Y.}~\bibnamefont {Xie}},\ }\href {\doibase
  10.1021/acssensors.7b00793} {\bibfield  {journal} {\bibinfo  {journal} {{ACS}
  Sensors}\ }\textbf {\bibinfo {volume} {3}},\ \bibinfo {pages} {167} (\bibinfo
  {year} {2017})}\BibitemShut {NoStop}%
\bibitem [{\citenamefont {Xiong}\ \emph {et~al.}(2019)\citenamefont {Xiong},
  \citenamefont {Zhang}, \citenamefont {Jiang}, \citenamefont {Yu},\ and\
  \citenamefont {Mao}}]{xiong_scc_2019}%
  \BibitemOpen
  \bibfield  {author} {\bibinfo {author} {\bibfnamefont {T.}~\bibnamefont
  {Xiong}}, \bibinfo {author} {\bibfnamefont {K.}~\bibnamefont {Zhang}},
  \bibinfo {author} {\bibfnamefont {Y.}~\bibnamefont {Jiang}}, \bibinfo
  {author} {\bibfnamefont {P.}~\bibnamefont {Yu}}, \ and\ \bibinfo {author}
  {\bibfnamefont {L.}~\bibnamefont {Mao}},\ }\href {\doibase
  10.1007/s11426-019-9526-4} {\bibfield  {journal} {\bibinfo  {journal} {Sci.
  China Chem.}\ }\textbf {\bibinfo {volume} {62}},\ \bibinfo {pages} {1346}
  (\bibinfo {year} {2019})}\BibitemShut {NoStop}%
\bibitem [{\citenamefont {Poggioli}, \citenamefont {Siria},\ and\ \citenamefont
  {Bocquet}(2019)}]{poggioli_jpcb_2019}%
  \BibitemOpen
  \bibfield  {author} {\bibinfo {author} {\bibfnamefont {A.~R.}\ \bibnamefont
  {Poggioli}}, \bibinfo {author} {\bibfnamefont {A.}~\bibnamefont {Siria}}, \
  and\ \bibinfo {author} {\bibfnamefont {L.}~\bibnamefont {Bocquet}},\ }\href
  {\doibase 10.1021/acs.jpcb.8b11202} {\bibfield  {journal} {\bibinfo
  {journal} {J. Phys. Chem. B}\ }\textbf {\bibinfo {volume} {123}},\ \bibinfo
  {pages} {1171} (\bibinfo {year} {2019})}\BibitemShut {NoStop}%
\bibitem [{\citenamefont {Cengio}\ and\ \citenamefont
  {Pagonabarraga}(2019)}]{dalcengio_jcp_2019}%
  \BibitemOpen
  \bibfield  {author} {\bibinfo {author} {\bibfnamefont {S.~D.}\ \bibnamefont
  {Cengio}}\ and\ \bibinfo {author} {\bibfnamefont {I.}~\bibnamefont
  {Pagonabarraga}},\ }\href {\doibase 10.1063/1.5108723} {\bibfield  {journal}
  {\bibinfo  {journal} {J. Chem. Phys.}\ }\textbf {\bibinfo {volume} {151}},\
  \bibinfo {pages} {044707} (\bibinfo {year} {2019})}\BibitemShut {NoStop}%
\bibitem [{\citenamefont {Kavokine}, \citenamefont {Netz},\ and\ \citenamefont
  {Bocquet}(2020)}]{kavokine_annualrev_2020}%
  \BibitemOpen
  \bibfield  {author} {\bibinfo {author} {\bibfnamefont {N.}~\bibnamefont
  {Kavokine}}, \bibinfo {author} {\bibfnamefont {R.~R.}\ \bibnamefont {Netz}},
  \ and\ \bibinfo {author} {\bibfnamefont {L.}~\bibnamefont {Bocquet}},\ }\href
  {\doibase 10.1146/annurev-fluid-071320-095958} {\bibfield  {journal}
  {\bibinfo  {journal} {Annu. Rev. Fluid Mech.}\ }\textbf {\bibinfo {volume}
  {53}} (\bibinfo {year} {2020}),\
  10.1146/annurev-fluid-071320-095958}\BibitemShut {NoStop}%
\bibitem [{\citenamefont {Noh}\ and\ \citenamefont
  {Aluru}(2020)}]{noh_acsnano_2020}%
  \BibitemOpen
  \bibfield  {author} {\bibinfo {author} {\bibfnamefont {Y.}~\bibnamefont
  {Noh}}\ and\ \bibinfo {author} {\bibfnamefont {N.~R.}\ \bibnamefont
  {Aluru}},\ }\href {\doibase 10.1021/acsnano.0c04453} {\bibfield  {journal}
  {\bibinfo  {journal} {{ACS} Nano}\ }\textbf {\bibinfo {volume} {14}},\
  \bibinfo {pages} {10518} (\bibinfo {year} {2020})}\BibitemShut {NoStop}%
\bibitem [{\citenamefont {Fertig}, \citenamefont {Valisk{\'{o}}},\ and\
  \citenamefont {Boda}(2020)}]{fertig_pccp_2020}%
  \BibitemOpen
  \bibfield  {author} {\bibinfo {author} {\bibfnamefont {D.}~\bibnamefont
  {Fertig}}, \bibinfo {author} {\bibfnamefont {M.}~\bibnamefont
  {Valisk{\'{o}}}}, \ and\ \bibinfo {author} {\bibfnamefont {D.}~\bibnamefont
  {Boda}},\ }\href {\doibase 10.1039/d0cp03237a} {\bibfield  {journal}
  {\bibinfo  {journal} {Phys. Chem. Chem. Phys.}\ }\textbf {\bibinfo {volume}
  {22}},\ \bibinfo {pages} {19033} (\bibinfo {year} {2020})}\BibitemShut
  {NoStop}%
\bibitem [{\citenamefont {Nernst}(1888)}]{nernst_zpc_1888}%
  \BibitemOpen
  \bibfield  {author} {\bibinfo {author} {\bibfnamefont {W.}~\bibnamefont
  {Nernst}},\ }\href {\doibase 10.1515/zpch-1888-0174} {\bibfield  {journal}
  {\bibinfo  {journal} {Zeitschrift für Physikalische Chemie}\ }\textbf
  {\bibinfo {volume} {2}} (\bibinfo {year} {1888}),\
  10.1515/zpch-1888-0174}\BibitemShut {NoStop}%
\bibitem [{\citenamefont {Planck}(1890)}]{planck_apc_1890}%
  \BibitemOpen
  \bibfield  {author} {\bibinfo {author} {\bibfnamefont {M.}~\bibnamefont
  {Planck}},\ }\href {\doibase 10.1002/andp.18902750202} {\bibfield  {journal}
  {\bibinfo  {journal} {Annalen der Physik und Chemie}\ }\textbf {\bibinfo
  {volume} {275}},\ \bibinfo {pages} {161} (\bibinfo {year}
  {1890})}\BibitemShut {NoStop}%
\bibitem [{\citenamefont {Boda}\ and\ \citenamefont
  {Gillespie}(2012)}]{boda_jctc_2012}%
  \BibitemOpen
  \bibfield  {author} {\bibinfo {author} {\bibfnamefont {D.}~\bibnamefont
  {Boda}}\ and\ \bibinfo {author} {\bibfnamefont {D.}~\bibnamefont
  {Gillespie}},\ }\href {\doibase 10.1021/ct2007988} {\bibfield  {journal}
  {\bibinfo  {journal} {J. Chem. Theor. Comput.}\ }\textbf {\bibinfo {volume}
  {8}},\ \bibinfo {pages} {824} (\bibinfo {year} {2012})}\BibitemShut {NoStop}%
\bibitem [{\citenamefont {Boda}\ \emph {et~al.}(2014)\citenamefont {Boda},
  \citenamefont {Kov\'acs}, \citenamefont {Gillespie},\ and\ \citenamefont
  {Krist\'of}}]{boda_jml_2014}%
  \BibitemOpen
  \bibfield  {author} {\bibinfo {author} {\bibfnamefont {D.}~\bibnamefont
  {Boda}}, \bibinfo {author} {\bibfnamefont {R.}~\bibnamefont {Kov\'acs}},
  \bibinfo {author} {\bibfnamefont {D.}~\bibnamefont {Gillespie}}, \ and\
  \bibinfo {author} {\bibfnamefont {T.}~\bibnamefont {Krist\'of}},\ }\href
  {\doibase 10.1016/j.molliq.2013.03.015} {\bibfield  {journal} {\bibinfo
  {journal} {J. Mol. Liq.}\ }\textbf {\bibinfo {volume} {189}},\ \bibinfo
  {pages} {100} (\bibinfo {year} {2014})}\BibitemShut {NoStop}%
\bibitem [{\citenamefont {Boda}(2014)}]{boda_arcc_2014}%
  \BibitemOpen
  \bibfield  {author} {\bibinfo {author} {\bibfnamefont {D.}~\bibnamefont
  {Boda}},\ }in\ \href {\doibase 10.1016/b978-0-444-63378-1.00005-7} {\emph
  {\bibinfo {booktitle} {Ann. Rep. Comp. Chem.}}},\ Vol.~\bibinfo {volume}
  {10},\ \bibinfo {editor} {edited by\ \bibinfo {editor} {\bibfnamefont
  {R.~A.}\ \bibnamefont {Wheeler}}}\ (\bibinfo  {publisher} {Elsevier},\
  \bibinfo {year} {2014})\ Chap.\ \bibinfo {chapter} {5 {Monte Carlo}
  Simulation of Electrolyte Solutions in Biology: {In} and Out of Equilibrium},
  pp.\ \bibinfo {pages} {127--163}\BibitemShut {NoStop}%
\bibitem [{\citenamefont {Fertig}\ \emph {et~al.}(2017)\citenamefont {Fertig},
  \citenamefont {M\'adai}, \citenamefont {Valisk\'o},\ and\ \citenamefont
  {Boda}}]{fertig_hjic_2017}%
  \BibitemOpen
  \bibfield  {author} {\bibinfo {author} {\bibfnamefont {D.}~\bibnamefont
  {Fertig}}, \bibinfo {author} {\bibfnamefont {E.}~\bibnamefont {M\'adai}},
  \bibinfo {author} {\bibfnamefont {M.}~\bibnamefont {Valisk\'o}}, \ and\
  \bibinfo {author} {\bibfnamefont {D.}~\bibnamefont {Boda}},\ }\href {\doibase
  10.1515/hjic-2017-0011} {\bibfield  {journal} {\bibinfo  {journal} {Hung. J.
  Ind. Chem.}\ }\textbf {\bibinfo {volume} {45}},\ \bibinfo {pages} {73}
  (\bibinfo {year} {2017})}\BibitemShut {NoStop}%
\bibitem [{\citenamefont {Gummel}(1964)}]{gummel1964self}%
  \BibitemOpen
  \bibfield  {author} {\bibinfo {author} {\bibfnamefont {H.~K.}\ \bibnamefont
  {Gummel}},\ }\href {\doibase 10.1109/t-ed.1964.15364} {\bibfield  {journal}
  {\bibinfo  {journal} {IEEE Transactions on electron devices}\ }\textbf
  {\bibinfo {volume} {11}},\ \bibinfo {pages} {455} (\bibinfo {year}
  {1964})}\BibitemShut {NoStop}%
\bibitem [{\citenamefont {Matejczyk}\ \emph {et~al.}(2017)\citenamefont
  {Matejczyk}, \citenamefont {Valisk{\'{o}}}, \citenamefont {Wolfram},
  \citenamefont {Pietschmann},\ and\ \citenamefont
  {Boda}}]{matejczyk_jcp_2017}%
  \BibitemOpen
  \bibfield  {author} {\bibinfo {author} {\bibfnamefont {B.}~\bibnamefont
  {Matejczyk}}, \bibinfo {author} {\bibfnamefont {M.}~\bibnamefont
  {Valisk{\'{o}}}}, \bibinfo {author} {\bibfnamefont {M.-T.}\ \bibnamefont
  {Wolfram}}, \bibinfo {author} {\bibfnamefont {J.-F.}\ \bibnamefont
  {Pietschmann}}, \ and\ \bibinfo {author} {\bibfnamefont {D.}~\bibnamefont
  {Boda}},\ }\href {\doibase 10.1063/1.4978942} {\bibfield  {journal} {\bibinfo
   {journal} {J. Chem. Phys.}\ }\textbf {\bibinfo {volume} {146}},\ \bibinfo
  {pages} {124125} (\bibinfo {year} {2017})}\BibitemShut {NoStop}%
\bibitem [{\citenamefont {M\'adai}\ \emph {et~al.}(2017)\citenamefont
  {M\'adai}, \citenamefont {Valisk\'o}, \citenamefont {Dallos},\ and\
  \citenamefont {Boda}}]{madai_jcp_2017}%
  \BibitemOpen
  \bibfield  {author} {\bibinfo {author} {\bibfnamefont {E.}~\bibnamefont
  {M\'adai}}, \bibinfo {author} {\bibfnamefont {M.}~\bibnamefont {Valisk\'o}},
  \bibinfo {author} {\bibfnamefont {A.}~\bibnamefont {Dallos}}, \ and\ \bibinfo
  {author} {\bibfnamefont {D.}~\bibnamefont {Boda}},\ }\href {\doibase
  10.1063/1.5007654} {\bibfield  {journal} {\bibinfo  {journal} {J. Chem.
  Phys.}\ }\textbf {\bibinfo {volume} {147}},\ \bibinfo {pages} {244702}
  (\bibinfo {year} {2017})}\BibitemShut {NoStop}%
\end{thebibliography}
% \bibliographystyle{unsrt} %the RSC's .bst file

\end{document}